\documentclass{emulateapj}

\newcommand{\logfxfr}{log($f_x/f_r$)}
\newcommand{\delpa}{$\Delta{\rm PA}$}

\slugcomment{To appear in AJ.}

\shorttitle{Radio Structure of HBLs}
\shortauthors{Rector et al.}

\begin{document}


\title{The Radio Structure of High-Energy Peaked BL Lacertae Objects}


\author{Travis A. Rector}
\affil{National Radio Astronomy Observatory, P.O. Box O, Socorro, NM 87801}
\email{trector@nrao.edu}
\author{Denise C. Gabuzda}
\affil{Department of Physics, National University of Ireland,  University College, Cork Ireland }

\and

\author{John T. Stocke}
\affil{Center for Astrophysics and Space Astronomy, University of Colorado, Boulder, CO 80309-0389}




\begin{abstract}

We present VLA and first-epoch VLBA observations that are part of a program to study the parsec-scale
radio structure of a sample of fifteen high-energy-peaked BL Lacs (HBLs).  The sample was chosen to
span the range of logarithmic X-ray to radio flux ratios observed in HBLs.  As this is only the first
epoch of observations, proper motions of jet components are not yet available; thus we consider only the
structure and alignment of the parsec- and kiloparsec-scale jets.  Like most low-energy-peaked BL
Lacs (LBLs), our HBL sample shows parsec-scale, core-jet morphologies and compact, complex
kiloparsec-scale morphologies.  Some objects also show evidence for bending of the jet 10--20pc from the
core, suggesting interaction of the jet with the surrounding medium.
Whereas LBLs show a wide distribution of parsec- to kpc-scale
jet misalignment angles, there is weak evidence that the jets in HBLs are more well-aligned, suggesting
that HBL jets are either intrinsically straighter or are seen further off-axis than LBL jets.


\end{abstract}




\section{Introduction}

BL Lacertae Objects are an extreme type of Active Galactic Nuclei (AGN), whose hallmark
is their ``featureless" optical spectrum.  By definition, any emission lines present must have rest
$W_{\lambda} \leq 5$\AA; e.g., \citet{sto91}.  BL Lacs are a member of the blazar class; and like
other blazars, they are characterized by their rapid variability, polarized optical and radio
emission and their  flat-spectrum radio emission.  See \citet{urr95} for an excellent summary of
their properties.  

%

Due to their luminous emission at these wavelengths, BL Lacs have been primarily discovered via X-ray
and radio surveys; thus, traditionally they have been labeled as radio-selected and X-ray-selected
(RBLs and XBLs respectively).   In recent years this terminology has given way to a more
physically meaningful classification based upon the overall spectral energy distribution (SED) of the
object.  In ``low-energy-peaked" blazars (LBLs), the peak of the synchrotron radiation occurs at
radio/IR wavelengths, whereas in ``high-energy-peaked" blazars (HBLs) this peak occurs in the UV/X-ray.
Historically, optical surveys have not been efficient in discovering BL Lacs
\citep{fle93,jan90}. It is therefore not surprising that, until recently, known BL Lacs have
shown a bimodal distribution of logarithmic X-ray to radio flux ratios, \logfxfr\ from 1~keV to 5~GHz,
with an approximate dividing line defined by \logfxfr\ $\sim -5.5$ \citep{wur96}.  For the most part
XBLs are HBLs and RBLs are LBLs, although exceptions do exist (e.g., Mkn 501 is an HBL in the 1Jy RBL
sample).  We choose to use the term HBL when referring to the BL Lacs studied herein because all were
chosen with a \logfxfr\ $\ga -5.5$ criterion.

New surveys, e.g., the ROSAT-Green Bank (RGB) survey \citep{lau99} and the DXRBS survey
\citep{per98}, have found BL Lacs with intermediate \logfxfr\ values, indicating a continuum of SEDs
in BL Lacs and likely rendering the LBL/HBL terminology obsolete.  The detailed properties of these
new BL Lac objects with intermediate \logfxfr\ values are not yet well-known.  So, although LBLs and
HBLs have different observed properties \citep[and references therein]{urr95}, it is not yet known
whether LBLs and HBLs are distinct classes of AGN or merely extremes of a continuum of properties for
a single class of AGN.  We have therefore chosen to include three intermediate objects from the RGB
sample in our sample of study.

Most BL Lacs are believed to be low-power radio galaxies whose jet axes are oriented at small angles to the
line of sight; and as such their jets are relativistically beamed, e.g., \citet{orr82,war84,urr95}. 
It has been proposed that HBLs and LBLs are both beamed, low-luminosity FR-I radio galaxies, and that
the observed differences between the two classes is a result of orientation.  In this unification model,
hereafter the ``orientation" model, HBLs are believed to be viewed further from the jet axis than LBLs
\citep{jan94,per94}.  The comparable X-ray luminosities for HBLs and LBLs require that either the
beam pattern or physical jet opening angle for the X-ray emission is larger than than for the radio
emission, e.g., \citet{sto89,urr91,ghi93,cel93}.  Several observed properties are consistent with this
hypothesis.  Compared to LBLs, HBLs are more numerous, their cores are less radio luminous, and they
are less optically variable \citep{jan93}.  HBLs also emit less compact radio emission
\citep{per93,per94,lau93}, contain larger fractions of starlight in the optical \citep{mor91}, and have
smaller degrees of optical polarization, often with a preferred optical polarization angle, $\pm
15\arcdeg$ \citep{jan94}.  However, several observed properties which should be independent of
orientation, e.g., observed evolution, optical emission-line strengths and extended radio powers, do not agree
between HBLs and LBLs, indicating that orientation alone is insufficient to explain the observed distinctions
between the two classes \citep{rec01}.  In fact, many LBLs appear to be beamed, high-luminosity FR-IIs; and at
least one is a gravitationally lensed object, the ``smallest Einstein ring" object 1Jy 0218+357 \citep{ode92}. 
The picture seems to be more clear for HBLs, whose properties are very consistent with being beamed FR-Is
\citep{rec00}.  Several observations support this picture, e.g., extended radio luminosity and morphology
\citep{ant85,per93}, host galaxy luminosity and morphology
\citep{abr91,wur96}, and comparative space densities and luminosity functions \citep{pad90,mor91}.

Alternatively, it has also been suggested that the difference between LBLs and HBLs lies
not in orientation, but in the high-energy cutoff in their energy distributions, such that
HBLs and LBLs represent a single family of objects with a smooth energy distribution
followed by a sharp cutoff.  For LBLs this cutoff occurs in the near-IR/optical and for
HBLs it is at UV/X-ray or higher energies \citep{gio94,pad95,sam96}.  In this model, hereafter the
``energy-cutoff" model, HBLs have intrinsically lower radio luminosities than LBLs, and
strong selection effects explain why most known BL Lacs are of the HBL variety.

Observational studies of BL Lacs at radio wavelengths have proven to be an effective test of
unification models for two reasons.  First, the kpc-scale extended radio flux is likely unbeamed, and
is therefore indicative of the actual AGN power; and second, the core and parsec-scale extended flux
is highly beamed, and is therefore strongly dependent on the orientation and relativistic nature of the
jet.  If HBLs are seen further off axis than LBLs, radio images will reveal several distinct trends. 
First, relativistic Doppler boosting and apparent proper motion are strongly sensitive to the
orientation angle and bulk Lorentz factor of the jet.  Thus, if seen close to their jet axis,
LBLs should be more core-dominated, show more instances of superluminal motion and have larger jet to
counter-jet brightness ratios.  And if HBLs and LBLs share the same parent population, both should have
similar jet Lorentz factors.  Second, geometrical projection effects will cause jets with intrinsically
small bends to appear highly distorted when seen close to the line of sight.  Thus, the projected
jet position angle (PA) in these highly inclined objects is very sensitive to the jet structure.  It
is known that LBLs show a wide range of parsec- and kpc-scale jet misalignment angles (\delpa\ $\equiv
\vert {\rm PA}_{kpc} - {\rm PA}_{pc} \vert$), presumably from a ``knotty" or helical jet seen close to
the line of sight 
\citep{kol92,con93,app96}.  If HBLs are seen further off-axis than LBLs, geometrical projection effects
will cause the parsec- and kpc-scale projected jet PAs of HBLs to be more aligned than LBLs.  


If the ``energy-cutoff" model is correct and there is no orientation bias, LBLs and HBLs should
show similar parsec- and kpc-scale radio structure, similar distributions of parsec- and kpc-scale
jet misalignment, assuming they share the same parent population.  But since LBLs are
generally more luminous than HBLs \citep{fos98}, their parsec-scale structures may differ
intrinsically from HBLs.  

In this paper we present deep VLA and first-epoch VLBA observations of a sample of fifteen BL Lacs which
span the full range of \logfxfr\ seen in HBLs (\S 2).   These maps are used to compare the parsec and
kpc-scale structure of these objects.  In
\S 3 we discuss the results of these observations, compare them to similar studies of LBLs and
discuss their implications for unification models.  As these are only the first epoch of
observations, proper motions of jet components are not yet available; thus we consider only the
alignment and structure of the parsec-scale jets.  In \S 4 we present the conclusions.  

\section{Observation and Reduction}

Due to their relative radio faintness, few HBLs have been studied with VLBI techniques. 
\citet{kol96} presented images of four HBLs from the bright HEAO 1 survey \citep{sch89}, all of which
have \logfxfr\ values close to the LBL/HBL boundary.  Conway et al. (private comm.) completed
VLBA observations of eight HBLs from the {\it Einstein} Medium Sensitivity Survey
\citep{mor91,rec00}; however, parsec-scale extended structure was not detected in most of these
objects, most likely due to their faintness ($f_{core} \approx 25$ mJy at 5~GHz).  

\begin{figure}
\plotone{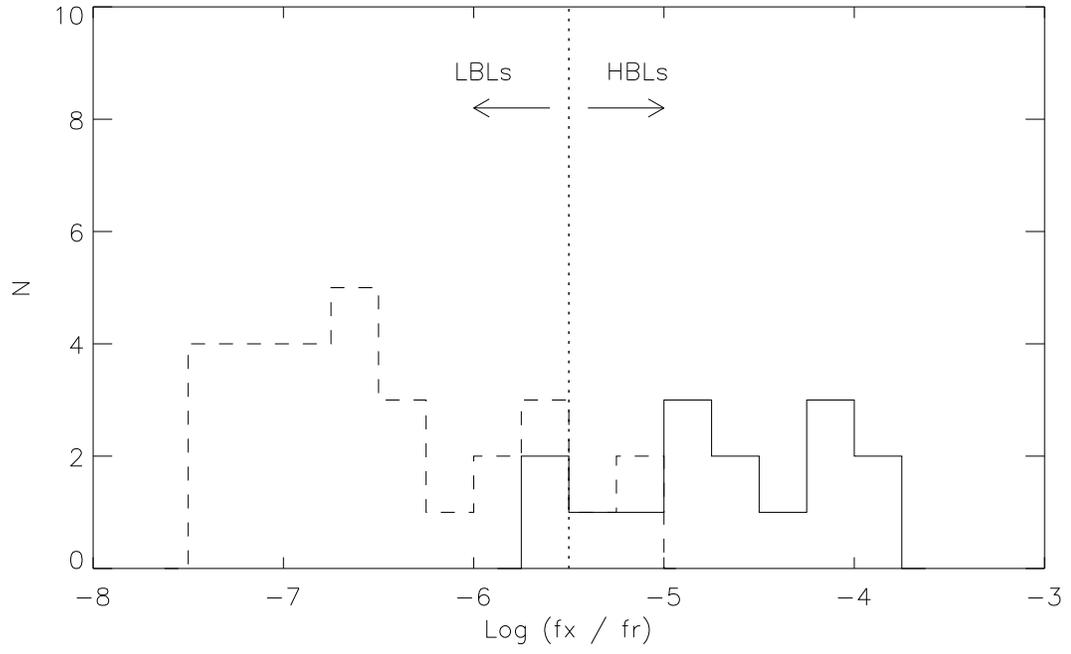}
\caption{Distribution of logarithmic X-ray to radio flux ratio \logfxfr\ for HBLs
in our sample (solid line) and LBLs in the 1Jy sample
\citep[dashed line]{sti91}.  The HBL/LBL dividing line at \logfxfr\ $\sim -5.5$ is shown.}
\label{fig-1}
\end{figure}

In order to better determine the parsec-scale structure of HBLs, we completed deep VLA and VLBA
observations of a sample of
HBLs that covers a factor of $\sim$30 in \logfxfr.  The initial goal was to determine their
morphology and core dominance on VLBI scales as well as to measure the projected alignment of their
parsec- and kpc-scale radio jets.   At the time this project was begun, the availability of
phase-reference calibrators for the VLBA was limited, so we chose only targets which were likely
bright enough for self-calibration ($f_{core} \geq 50$ mJy at 4.964 GHz).  Since HBLs are for the most
part too radio faint to observe a statistically complete sample without phase referencing, we chose a
sample that represented the full range of \logfxfr\ values observable in HBLs, as shown in
Figure~\ref{fig-1}. Our sample was selected primarily from the {\it Einstein} ``Slew" Survey (1ES;
\citet{per96}).  Objects that are LBL-like (i.e., \logfxfr\ $< -5.5$) were discarded.  This sample was
supplemented with three objects from the RGB BL Lac sample
to fill in the \logfxfr\ distribution with objects intermediate to LBLs and HBLs.  The final sample
consists of fifteen HBLs with an even distribution of
\logfxfr\ ranging from $-3.9$6 to $-5.66$, and is compared to the complete 1Jy LBL sample in
Figure~\ref{fig-1}.  

A summary of the sample properties is given in Table~\ref{tbl-1}.  The columns are: [1] the object
name; [2-3] the Right Ascension and Declination in J2000 coordinates; [4] the
\logfxfr\ value \citep{per96,lau99}; and [5] the redshift \citep{lau99,per99}.   The redshifts for 1ES
0647+250 and 1ES 1028+511 are marked with a colon because they are tentative.

\subsection{VLBA Observations}

Twelve BL Lacs in our sample were observed with the NRAO\footnote{The National Radio Astronomy
Observatory is operated by Associated Universities, Inc., under cooperative agreement with the National
Science Foundation.} Very Long Baseline Array (VLBA) on 17 May 1997.   We chose to observe at 4.964 GHz in
order to obtain both high resolution and high sensitivity to maximize the probability of detecting and
resolving faint VLBI jets in these compact objects.  A single 24-hour time allocation was
used with the complete VLBA array of 10 x 25m antennas.  This allowed approximately twelve 6.5-minute
scans per source, which were well-spaced in hour angle for optimal ($u,v$) plane coverage.  To minimize
potential phase and amplitude errors due to poor source positions in the correlator model,
high-resolution ``snapshots" were completed with the NRAO Very Large Array (VLA) B-array at 3.6cm prior
to the VLBA observations to obtain core positions with 0.01\arcsec\ accuracy.  These are the positions 
given in Table~\ref{tbl-1}.  The data were initially calibrated with the AIPS software package in the
standard manner.  Self-calibration and imaging were then done with the DIFMAP software package
\citep{shep97}.  The resultant maps each have an RMS noise level of $\sigma \approx 0.1$ mJy
beam$^{-1}$.  The three RGB sources in our sample were observed by \citet{bon01} and \citet{fey00}.

All of the objects except for 1ES 0647+250 and 1ES 1028+511 are clearly resolved; however these two
objects do show evidence for very faint extended structure.  In all sources 60\% to 90\% of the VLA
core flux, as determined by snapshots in \citet{per96}, was detected on parsec scales.  All of the
resolved objects show a core-jet morphology, as is typical of LBLs and core-dominated quasars
\citep{pea88}.  All are core dominated on parsec scales ($R \equiv f_{core}/f_{ext} \geq 1$).
Some sources show a well-collimated jet with
discrete components (e.g., 1ES 1741+196 and 1ES 1212+078), whereas others show a diffuse jet with
a wide opening angle ($> 60$\arcdeg; e.g., 1ES 0806+524 and 1ES 1959+650).  A map of 1E 1415+259 is
not shown because its extended structure is too faint to be imaged.  The peak component
flux is of the order of the baseline noise ($\sim$7 mJy), and baseline phase errors dominate the
apparent morphology of the extended structure.  

\subsection{VLA Observations}

Eleven BL Lacs in our sample were observed with the VLA on 10 October 1998.   We
chose to observe with the B-array at 1.425~GHz with a 50MHz bandwidth to maximize sensitivity to extended,
steep-spectrum structure while achieving
$\sim$4\arcsec\ resolution.  The B-array was chosen based upon the redshifts of the objects in our sample to
avoid over-resolution of the extended structure.  Three or four 8-minute scans, each bracketed by a 90--second
scan on a primary VLA flux calibrator, were made for each source.  Scans were spaced to optimize coverage in the
$(u,v)$ plane.  Objects which were sufficiently resolved in previous efforts
\citep{lau93,per96} were not reobserved here.

Epoch 1995.2 VLA values were used to flux calibrate the maps using multiple observations of 3C 286. 
Since these sources are highly core dominated, a point source model was assumed to start the
self-calibration process.  Phase-only self-calibration in
decreasing solution time intervals was used for the first four iterations.  Amplitude and phase
self-calibration were then used until the maximum dynamic range was achieved, usually requiring only
one or two more iterations.  The AIPS task IMAGR was used to generate the maps and clean components. 
Robust weighting (ROBUST = 0.5) was used to achieve a smaller beam FWHM with only a 10--12\% increase
in noise over natural weighting; see \citet{bri95} for an explanation.
The core flux densities were measured by fitting the core with a single Gaussian with the synthesized
beam's parameters.  The extended flux was determined by measuring the total flux density with a box
enclosing the entire source and then subtracting the core flux density.  
For unresolved sources, conservative
upper limits on extended radio flux densities were calculated by assuming that each
source has uniformly bright extended emission at the 1$\sigma$ detection level over
a 3000 kpc$^2$ area surrounding the core.

\subsection{Summary of Radio Properties}

\begin{figure}
\plotone{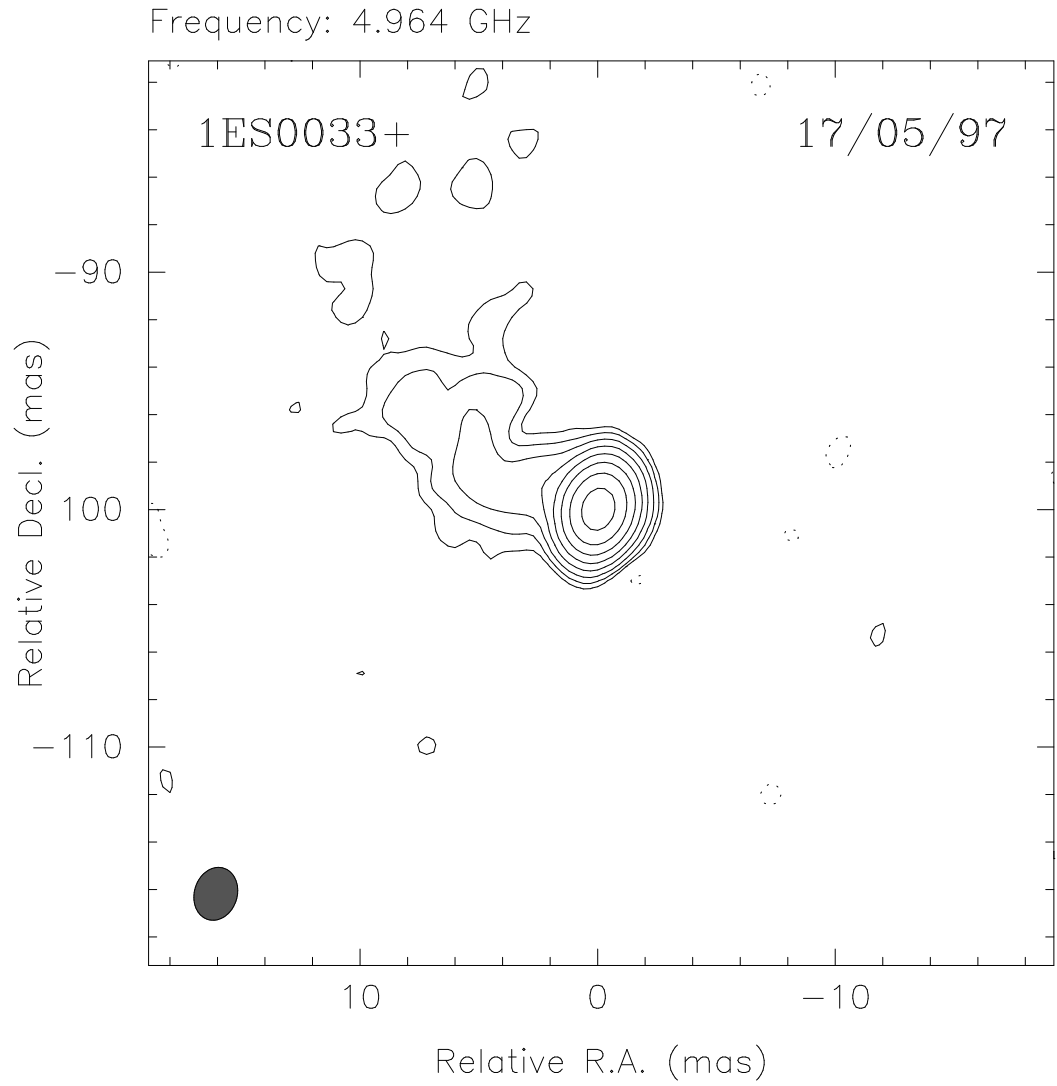}
\caption{VLBA 4.964 GHz map of 1ES 0033+595.  The contour levels are $-0.54$, 0.54, 1.07, 2.15, 4.30, 8.59,
17.18, 34.37 and 68.73\% of the peak flux of 3.72 x 10$^{-2}$ Jy beam$^{-1}$.  The beam, shown in the lower
left corner, has a FWHM of 2.26 x 1.80 mas, PA $-17.0$\arcdeg.   The date of observation is shown in the upper
right corner of the map.}
\label{fig-4}
\end{figure}

\begin{figure}
\plotone{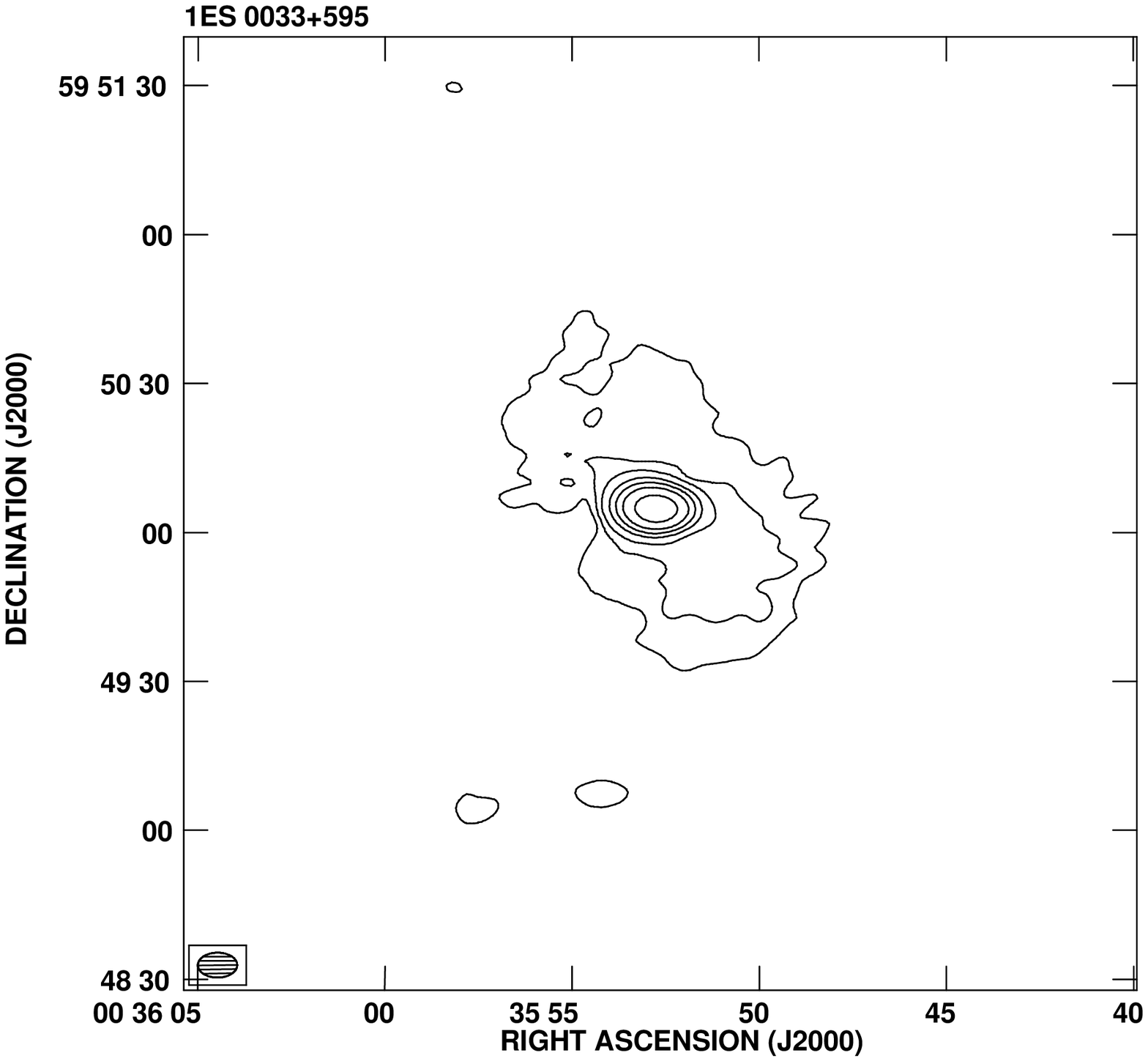}
\caption {VLA 1.425 GHz map of 1ES 0033+595.  The beam is shown in the lower left corner.  The contour levels
are 0.5, 1, 2, 5, 10, 20, 50 and 100\% the peak flux of 8.39 x 10$^{-2}$ Jy
beam$^{-1}$.}
\label{fig-5}
\end{figure}

\begin{figure}
\plotone{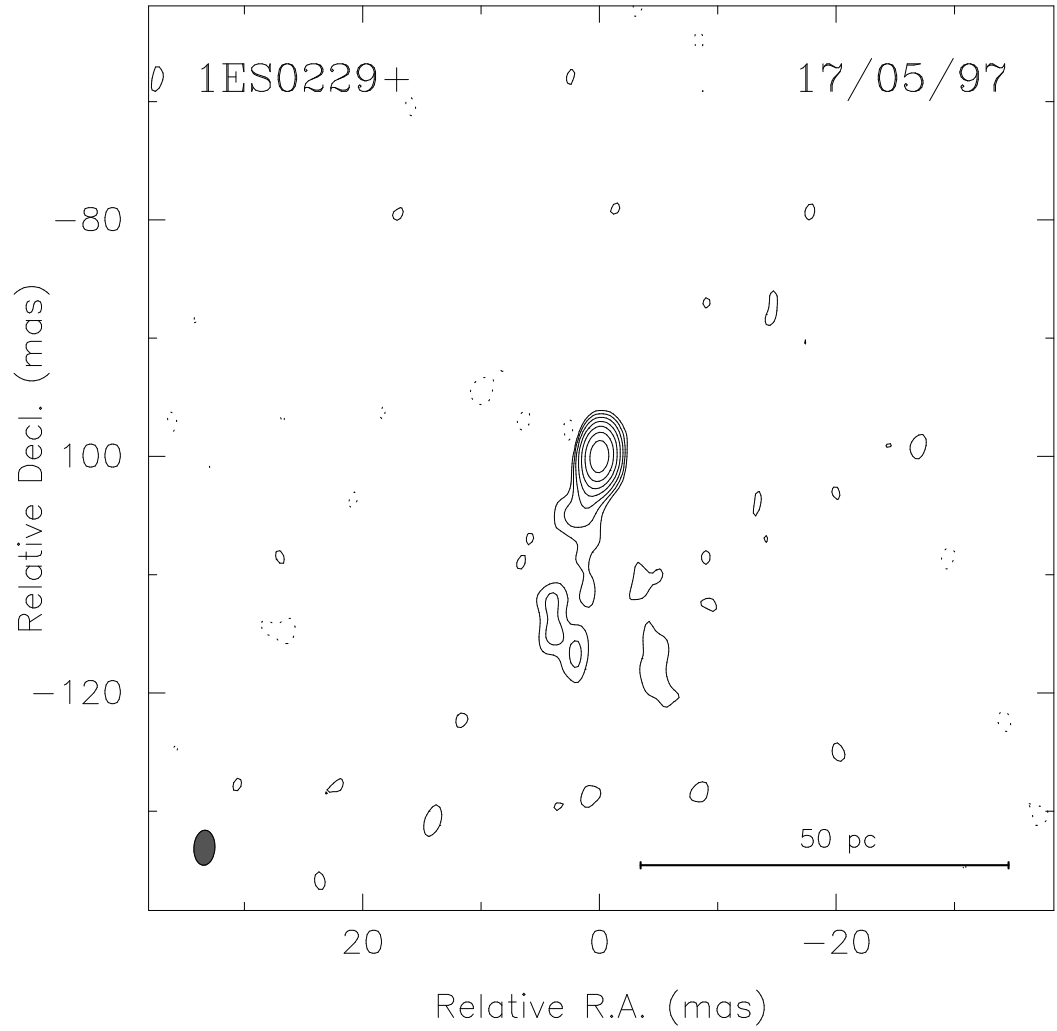}
\caption{VLBA 4.964 GHz map of 1ES 0229+200.  The contour levels are $-0.91$, 0.91, 1.82, 3.63, 7.27, 14.54,
29.07 and 58.15\% of the peak flux of 2.20 x 10$^{-2}$ Jy beam$^{-1}$.  The beam, shown in the lower left
corner, has a FWHM of 2.97 x 1.78 mas, PA $-3.3$\arcdeg.   The date of observation is shown in the upper
right corner of the map.}
\label{fig-6}
\end{figure}

\begin{figure}
\plotone{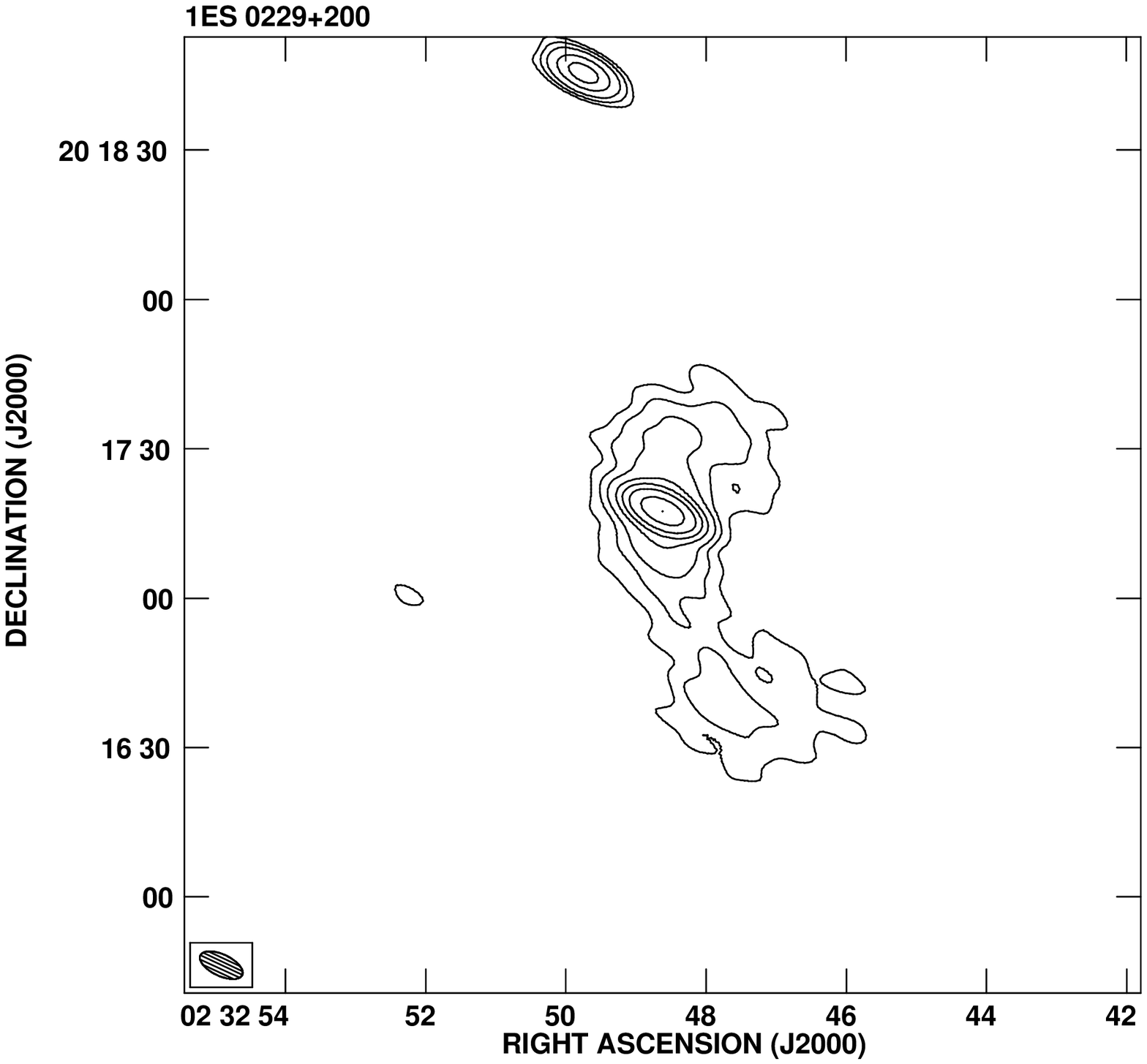}
\caption{VLA 1.425 GHz map of 1ES 0229+200.  The beam is shown in the lower left corner.  The contour levels
are 0.5, 1, 2, 5, 10, 20, 50 and 100\% the peak flux of 4.99 x 10$^{-2}$ Jy
beam$^{-1}$.}
\label{fig-7}
\end{figure}

\begin{figure}
\plotone{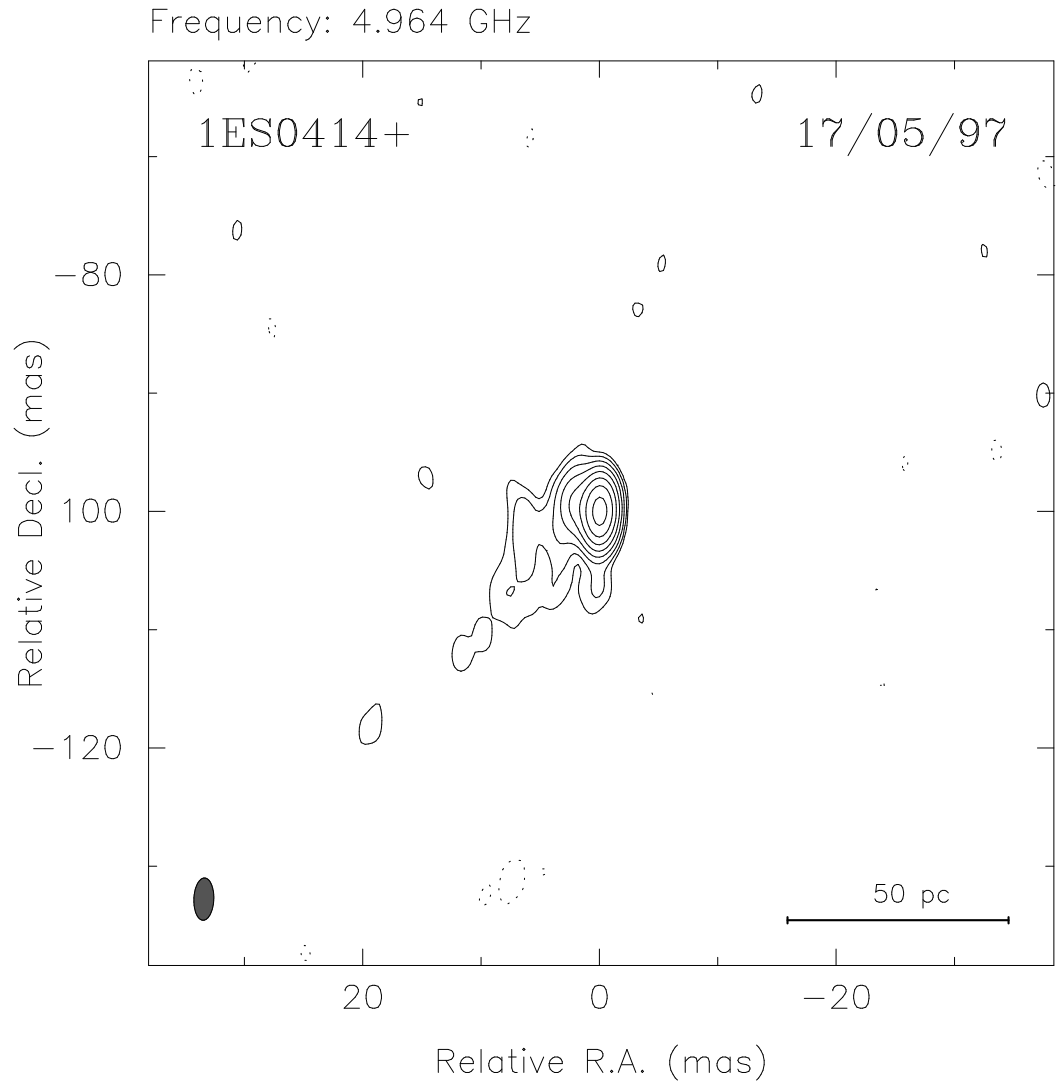}
\caption{VLBA 4.964 GHz map of 1ES 0414+009.  The contour levels are $-0.58$, 0.58, 1.15, 2.31, 4.62, 9.23,
18.47, 36.93 and 73.86\% of the peak flux of 3.47 x 10$^{-2}$ Jy beam$^{-1}$.  The beam, shown in the lower
left corner, has a FWHM of 3.58 x 1.68 mas, PA $-2.2$\arcdeg.   The date of observation is shown in the upper
right corner of the map.}
\label{fig-8}
\end{figure}

\begin{figure}
\plotone{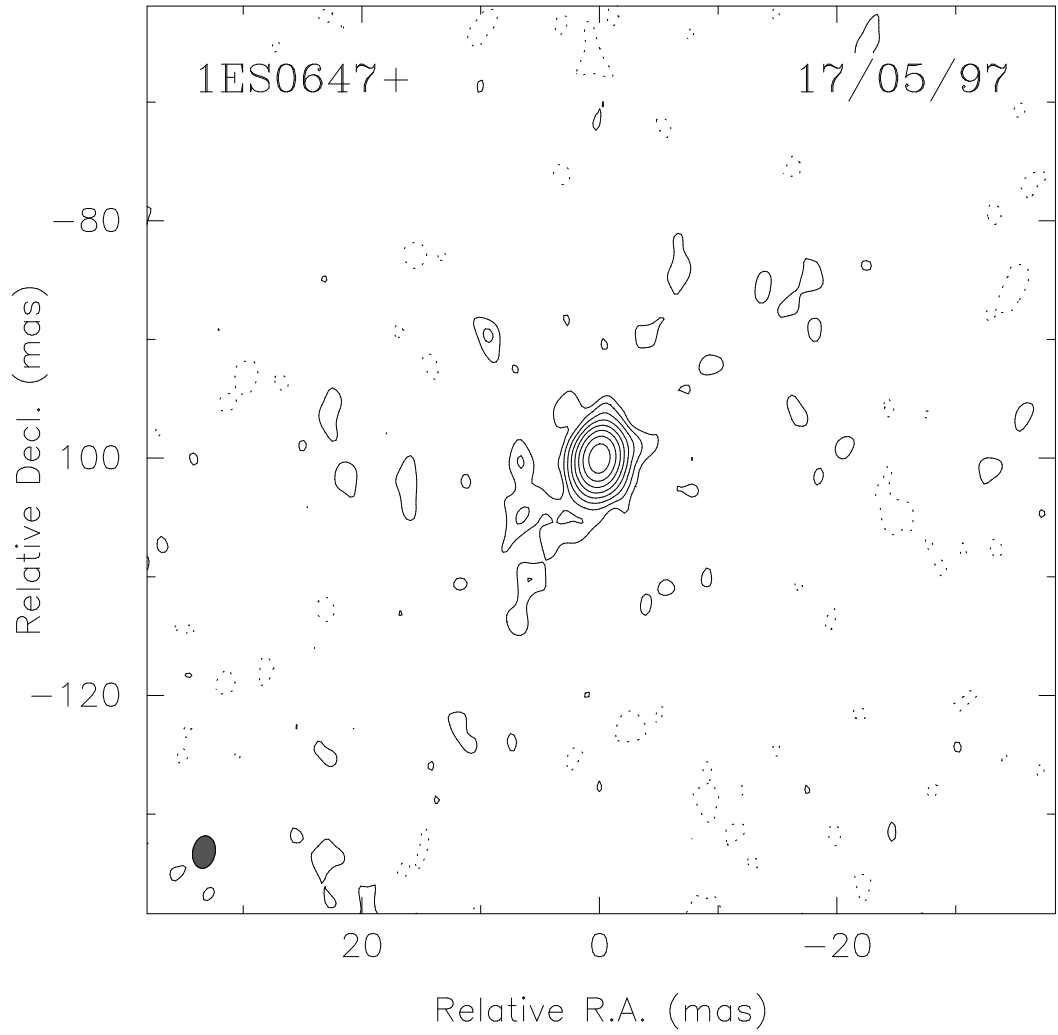}
\caption{VLBA 4.964 GHz map of 1ES 0647+250.  The contour levels are $-0.46$, 0.46, 0.92, 1.84, 3.68, 7.36,
14.72, 29.44 and 58.88\% of the peak flux of 4.35 x 10$^{-2}$ Jy beam$^{-1}$.  The beam, shown in the lower
left corner, has a FWHM of 2.76 x 1.90 mas, PA $-10.3$\arcdeg.   The date of observation is shown in the upper
right corner of the map.}
\label{fig-9}
\end{figure}

\begin{figure}
\plotone{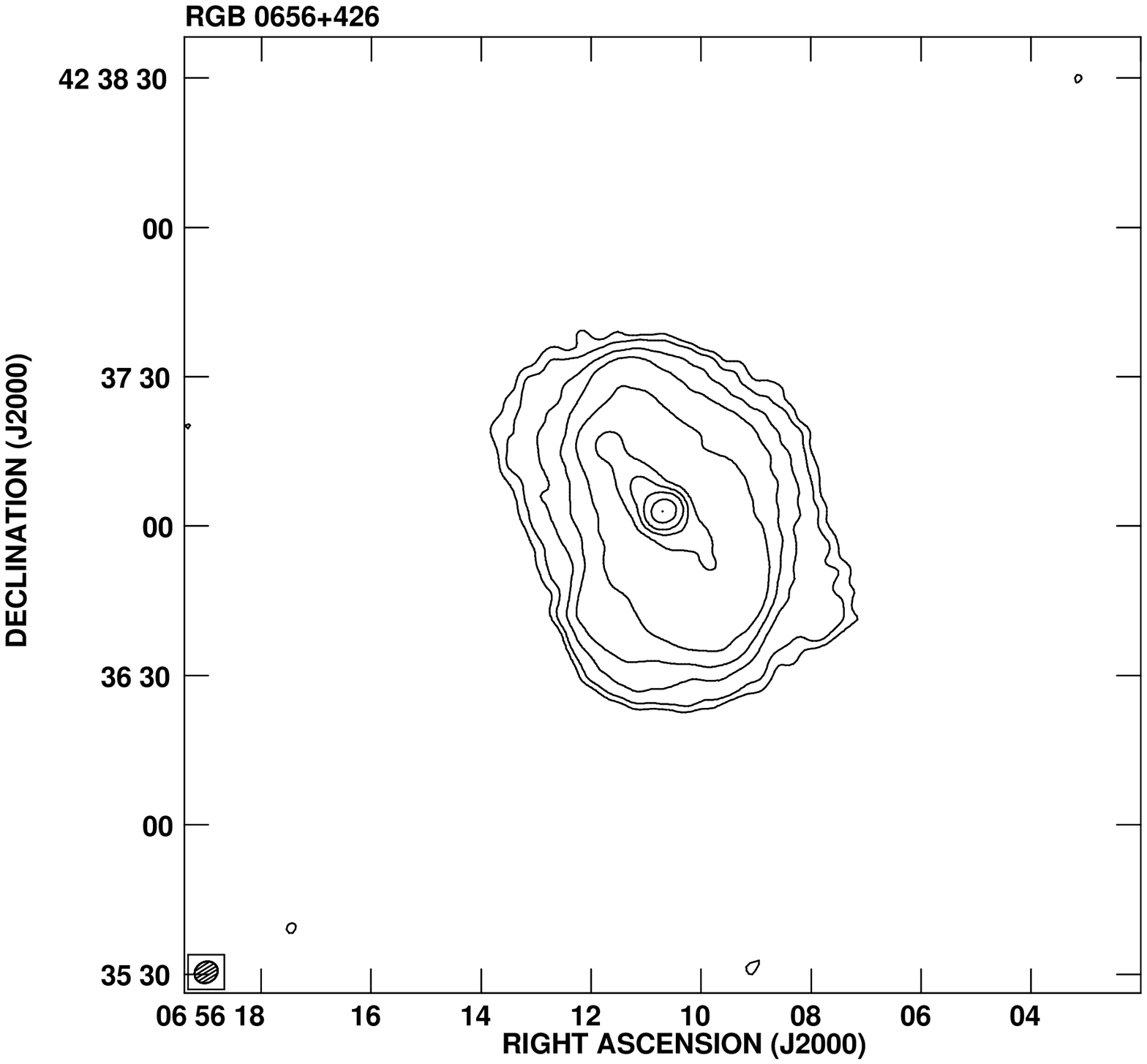}
\caption{VLA 1.425 GHz map of RGB 0656+426.  The beam is shown in the lower left corner.  The contour levels
are 0.1, 0.2, 0.5, 1, 2, 5, 10, 20, 50 and 100\% the peak flux of 2.308 x 10$^{-1}$ Jy beam$^{-1}$.}
\label{fig-10}
\end{figure}

\begin{figure}
\plotone{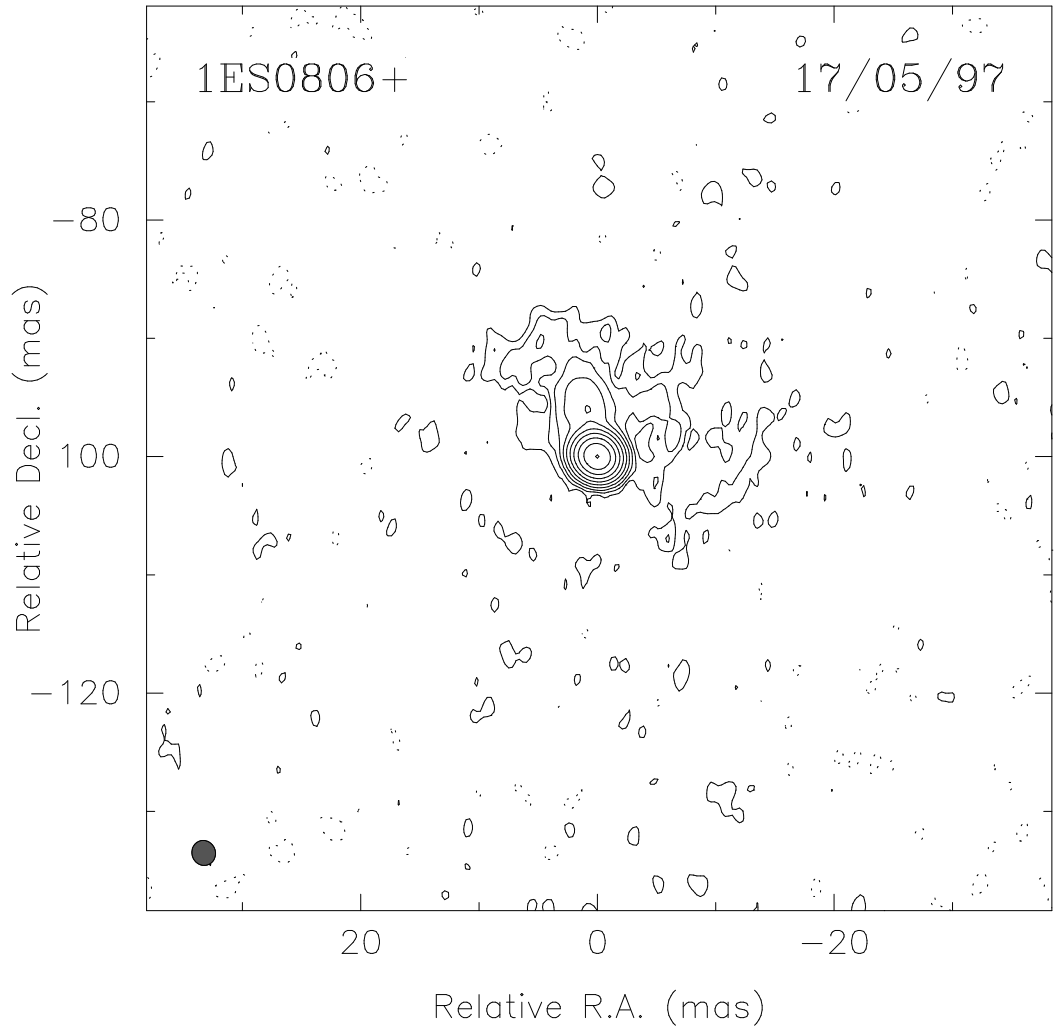}
\caption{VLBA 4.964 GHz map of 1ES 0806+524.  The contour levels are $-0.19$, 0.19, 0.38, 0.76, 1.53, 3.05,
6.11, 12.22, 24.43, 48.86 and 97.73\% of the peak flux of 1.048 x 10$^{-1}$ Jy beam$^{-1}$.  The beam, shown
in the lower left corner, has a FWHM of 2.13 x 1.99 mas, PA
$18.9$\arcdeg.   The date of observation is shown in the upper right corner of the map.}
\label{fig-11}
\end{figure}

\begin{figure}
\plotone{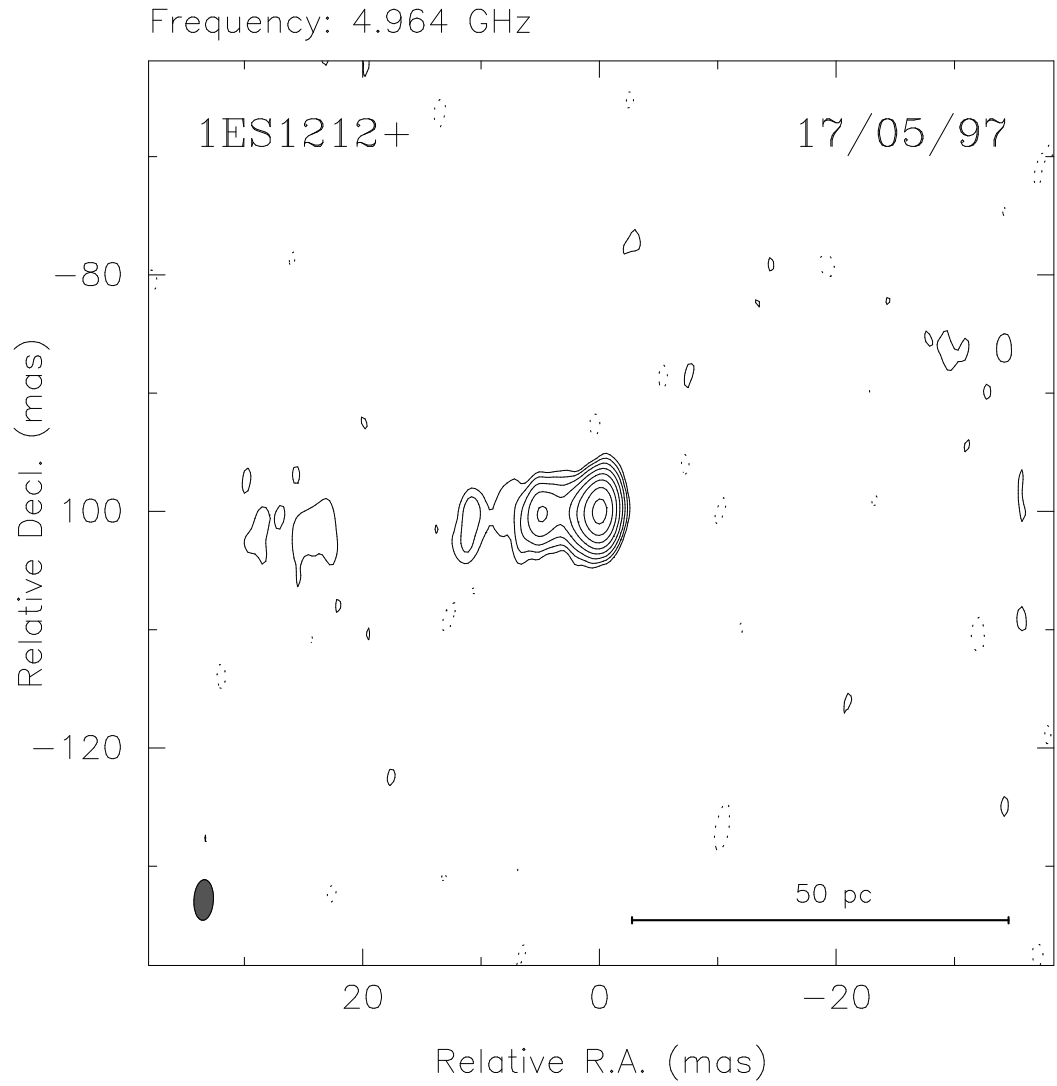}
\caption{VLBA 4.964 GHz map of 1ES 1212+078.  The contour levels are $-0.60$, 0.60, 1.20, 2.40, 4.80, 9.59,
19.18, 38.36 and 76.72\% of the peak flux of 3.33 x 10$^{-2}$ Jy beam$^{-1}$.  The beam, shown in the lower
left corner, has a FWHM of 3.44 x 1.65 mas, PA
$-3.2$\arcdeg.   The date of observation is shown in the upper right corner of the map.}
\label{fig-12}
\end{figure}

\begin{figure}
\plotone{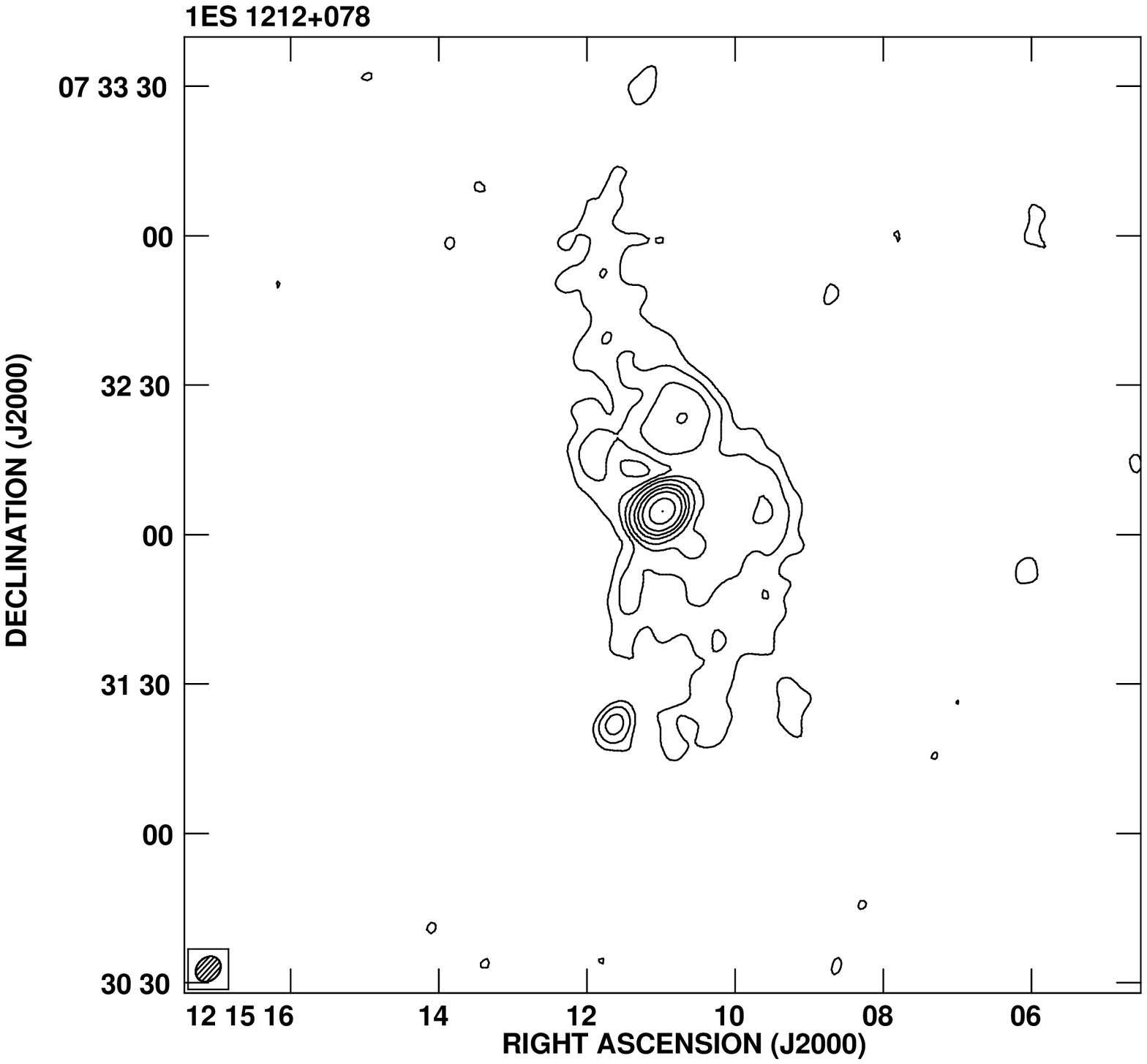}
\caption{VLA 1.425 GHz map of 1ES 1212+078.  The beam is shown in the lower left corner.  The contour levels
are 0.2, 0.5, 1, 2, 5, 10, 20, 50 and 100\% the peak flux of 8.60 x 10$^{-2}$ Jy beam$^{-1}$.}
\label{fig-13}
\end{figure}

\begin{figure}
\plotone{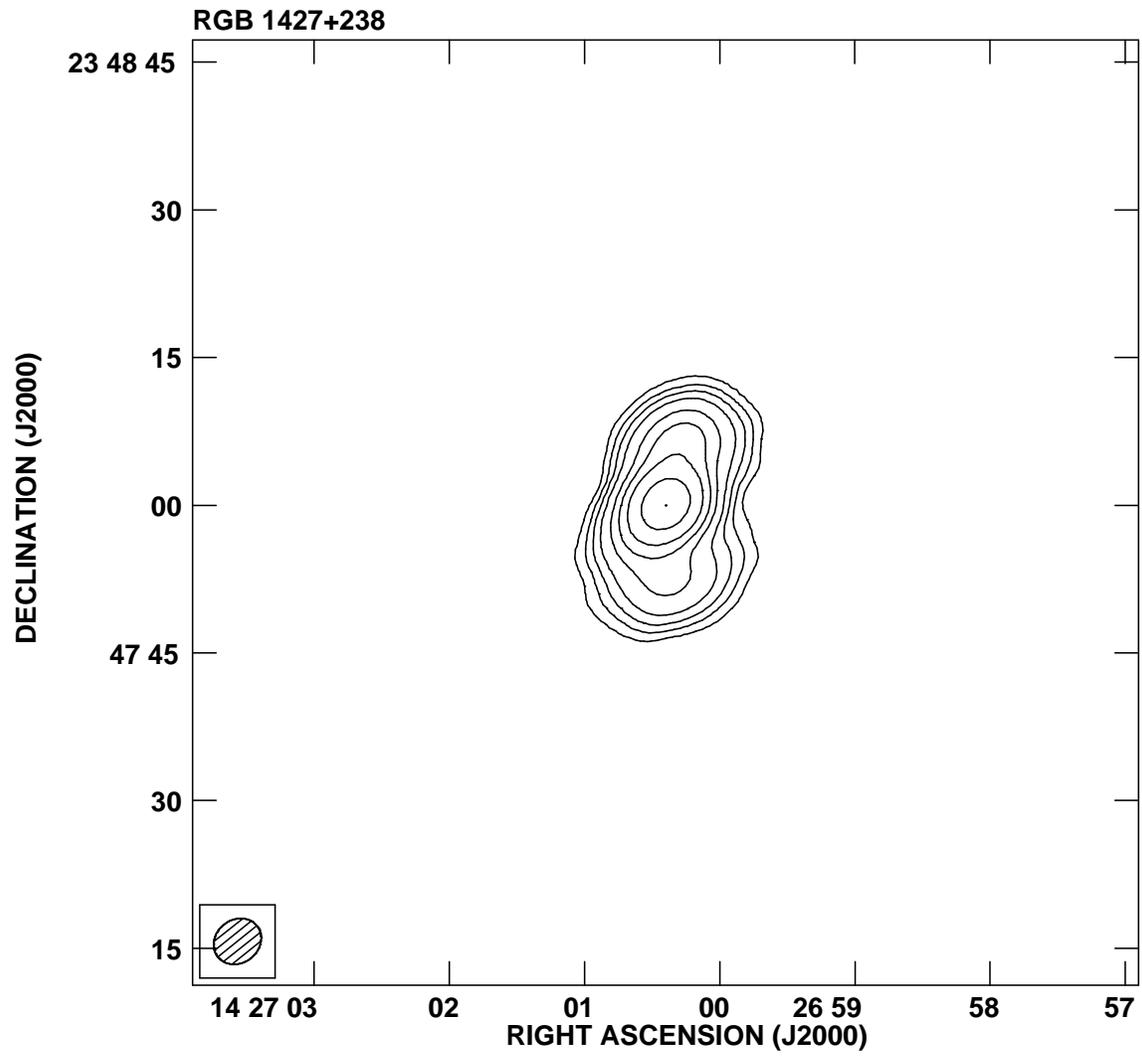}
\caption{VLA 1.425 GHz map of RGB 1427+238.  The beam is shown in the lower left corner.  The contour levels
are 0.2, 0.5, 1, 2, 5, 10, 20, 50 and 100\% the peak flux of 3.102 x 10$^{-1}$ Jy beam$^{-1}$.}
\label{fig-14}
\end{figure}

\begin{figure}
\plotone{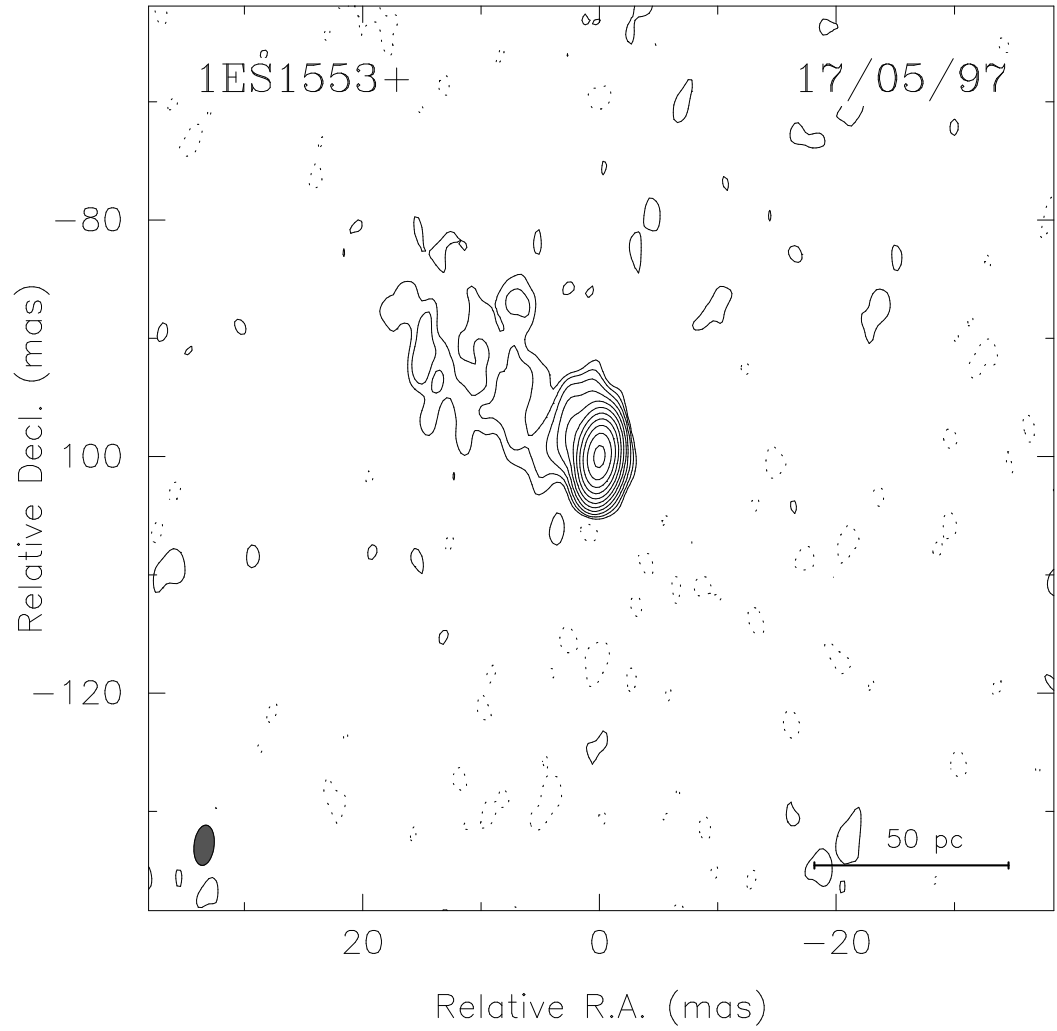}
\caption{VLBA 4.964 GHz map of 1ES 1553+113.  The contour levels are $-0.08$, 0.08, 0.16, 0.32, 0.63, 1.27,
2.53, 5.06, 10.12, 20.24, 40.49 and 80.98\% of the peak flux of 2.529 x 10$^{-1}$ Jy beam$^{-1}$.  The beam,
shown in the lower left corner, has a FWHM of 3.42 x 1.70 mas, PA
$-6.7$\arcdeg.   The date of observation is shown in the upper right corner of the map.}
\label{fig-15}
\end{figure}

\begin{figure}
\plotone{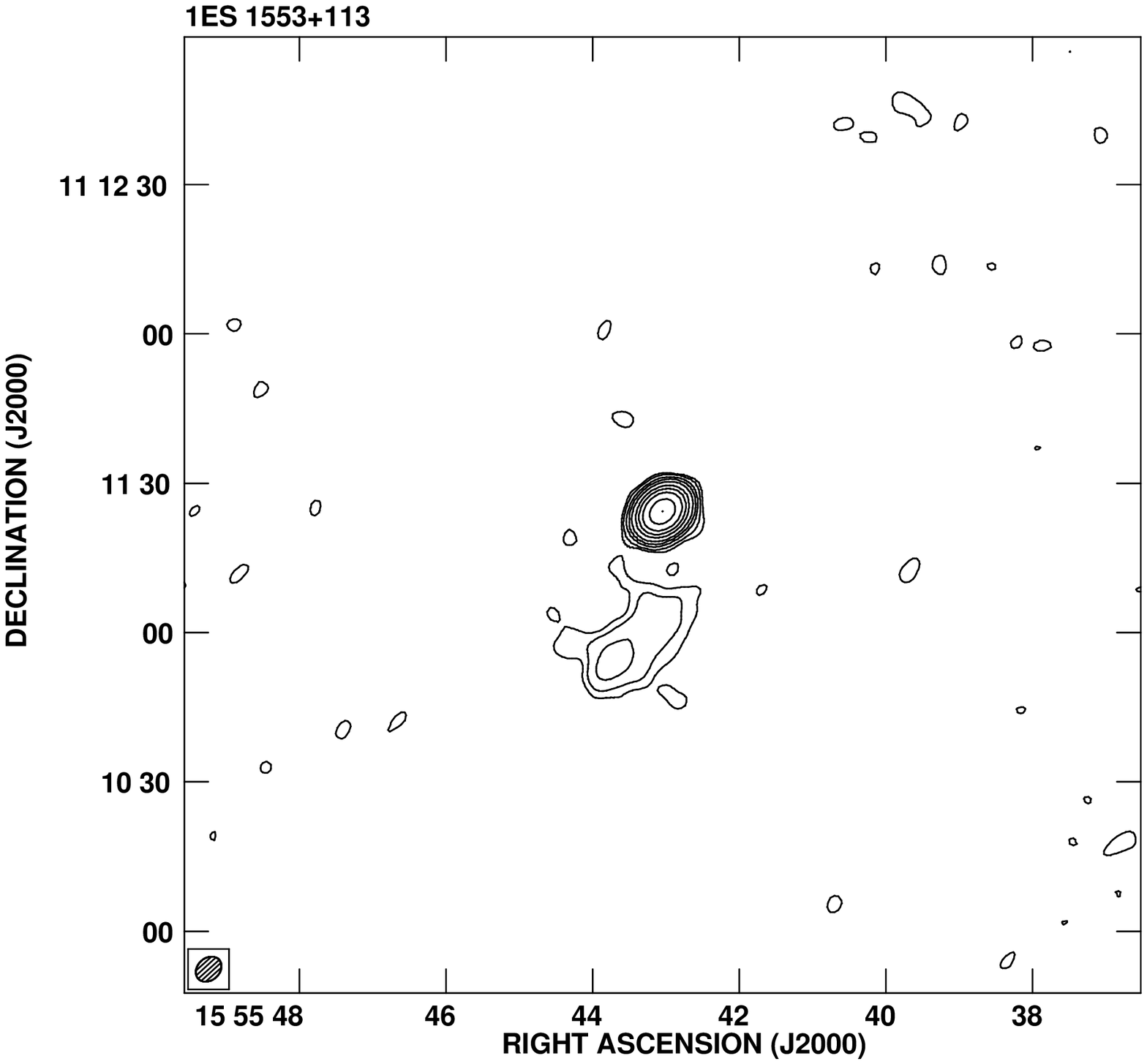}
\caption{VLA 1.425 GHz map of 1ES 1553+113.  The beam is shown in the lower left corner.  The contour levels
are 0.1, 0.2, 0.5, 1, 2, 5, 10, 20, 50 and 100\% the peak flux of 2.706 x 10$^{-1}$ Jy beam$^{-1}$.}
\label{fig-16}
\end{figure}

\begin{figure}
\plotone{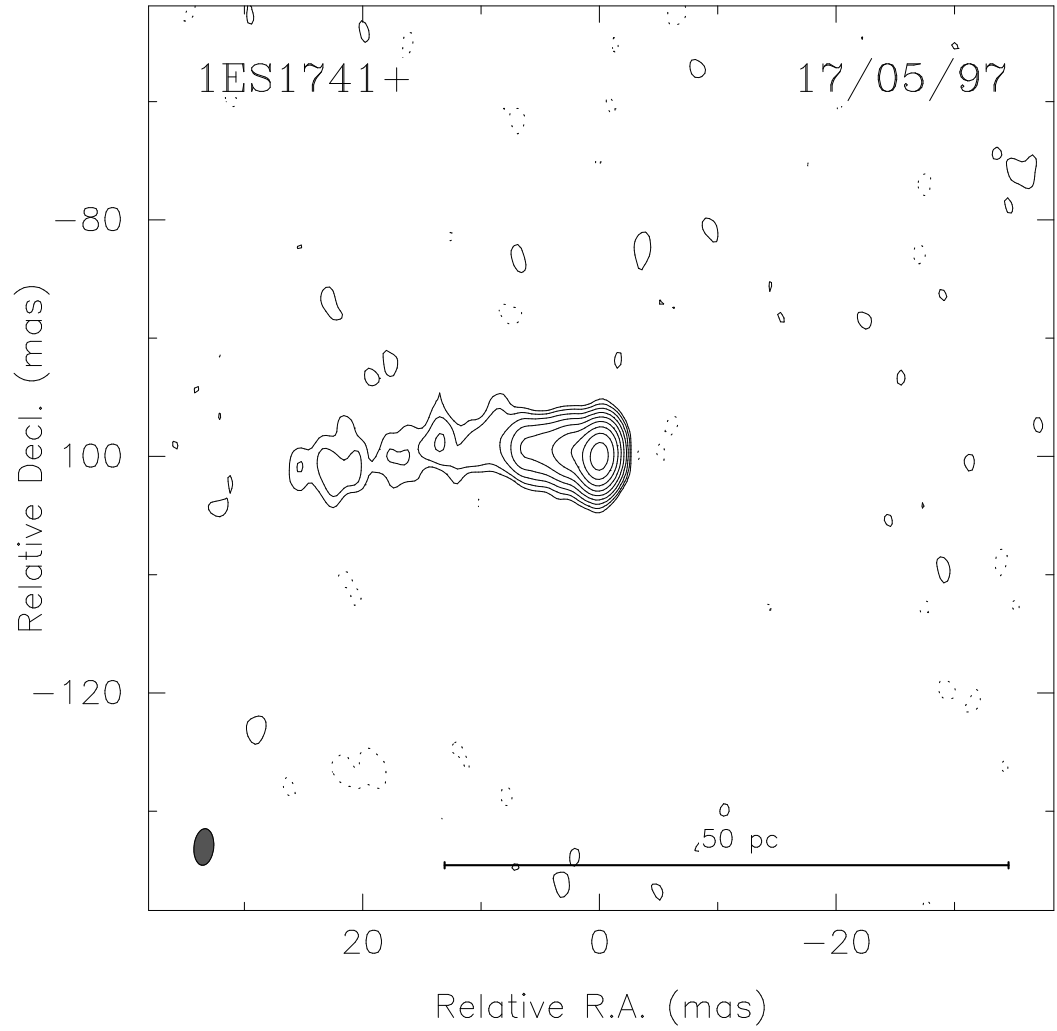}
\caption{VLBA 4.964 GHz map of 1ES 1741+196.  The contour levels are $-0.26$, 0.26, 0.53, 1.05, 2.11, 4.21,
8.42, 16.84, 33.69 and 67.38\% of the peak flux of 7.60 x 10$^{-2}$ Jy beam$^{-1}$.  The beam, shown in the
lower left corner, has a FWHM of 3.10 x 1.67 mas, PA
$-5.4$\arcdeg.   The date of observation is shown in the upper right corner of the map.}
\label{fig-17}
\end{figure}

The VLA and VLBA maps are shown in Figures~\ref{fig-4} through~\ref{fig-22}.  A summary of the radio
properties is given in Table~\ref{tbl-2}.  The columns are: [1] the object name; [2-3] the 1.425~GHz VLA
core and extended flux densities (mJy); [4] the VLA jet position angle
[5-6] the 4.964 GHz VLBA core and extended flux densities (mJy); and [7] the VLBA jet position angle.
The errors in the flux densities are $\sim$0.1 mJy beam$^{-1}$ for both the VLA and VLBA maps.  The cumulative
errors in the extended flux densities depend upon the solid angular extent of the measured flux.  Note that, as
discussed in \S 3, measured
jet position angles are subjective; and in most objects the jet morphology is more complex than what can be
modeled with a single PA value.  Thus the discussion of individual sources below should be consulted for each
source.

\subsection{Discussion of Individual Sources}

1ES 0033+595:  The VLBA map (Figure~\ref{fig-4}) shows a diffuse jet extending in PA = +65\arcdeg.  It
is not clear whether or not the parsec-scale jet is well-collimated.  There is some evidence for the
jet curving to the north.  Thus the measured PA$_{\rm kpc}$ is likely resolution dependent.  The VLA image
(Figure~\ref{fig-5}) of this source shows a diffuse  halo surrounding the core.  Jets are not clearly resolved;
however the core is elongated in the PA = $+62$\arcdeg\ and PA = $-105$\arcdeg\ directions, so the parsec- and
kpc-scale jets appear to be well aligned if the parsec-scale jet is assumed to be related with the PA =
$+62$\arcdeg\ extension of the VLA core.  {\it HST} PC observations of this object shows two unresolved sources
of similar brightness separated by 1.58\arcsec\ \citep{sca99}, which might have been explained as multiple images
of a gravitationally lensed system.  However, the VLA astrometric observations presented here do not detect a
second radio source, ruling out that possibility.  

1ES 0229+200:  The VLBA map (Figure~\ref{fig-6}) shows a jet extending to the south (PA =
+170\arcdeg), with weak evidence for a broad jet opening angle of $\sim$30\arcdeg.  The VLA image
(Figure~\ref{fig-7}) shows curved jets to the north and south, both of which curve to the west. 
Measuring the jet PAs from the the inner contours gives PA $= -10$\arcdeg\ and PA = 180\arcdeg.  The
VLBA jet appears to be well-aligned with the southern jet (\delpa\ = $10$\arcdeg).    

1ES 0414+009:  The VLBA map (Figure~\ref{fig-8}) resolves a jet which initially extends to the
east-northeast (PA = $+68$\arcdeg) of the core.   There is also weak, extended ($\sim$3-4$\sigma$)
emission to the southeast of the jet, which suggests either that the jet is collimated and bends to
the south $\sim$10pc from the core, or that the projected jet opening angle is wide
($\sim$60\arcdeg).  If the jet does indeed bend to the southeast, the misalignment angle could be as
large as \delpa\ = $\sim$60\arcdeg; however the inner portion of the jet  is well-aligned (\delpa\
$=5$\arcdeg) to the kpc-scale jet shown in Figure 3 of \citet{lau93}.

1ES 0647+250:  The VLBA map (Figure~\ref{fig-9}) of this source does not show a distinct jet but there
is some evidence for a faint, diffuse halo around the core, with weak evidence of elongation of
the core to the north at PA = $-10$\arcdeg.  VLA snapshot observations \citep{per93} of this source show
a jet extending to the southwest (PA = $-124$\arcdeg) as well as a possible extention to the northwest;
although the reality of the extended structure is seriously questioned due to the poor quality of this
map.  A deep VLA map is necessary to better determine the kpc-scale structure of this source.  Due to
the uncertainty of the VLA and VLBA structure a \delpa\ for this source is not considered in analysis.  The
redshift of 1ES 0647+250 is tentative and must be confirmed by further optical spectroscopy.

RGB 0656+426:  The VLA map of this source (Figure~\ref{fig-10}) shows a jet-core-jet source
embedded in a bright halo, with a jet at PA $= +40$\arcdeg\ and a colinear
counter-jet.  The VLBA map of this source \citep{bon01} resolves a well-collimated jet at
PA $= -150$\arcdeg, which is well-aligned (\delpa\ $=10$\arcdeg) with the counter-jet. 
The halo morphology on kpc scales suggests this object is seen very close to the jet axis.

1ES 0806+524:  The VLBA map (Figure~\ref{fig-11}) shows a jet to the north with PA = $+13$\arcdeg. 
Very faint, diffuse extended emission surrounds the jet, suggesting it has a broad opening angle
which may be as wide as 70\arcdeg.  A deep VLA map of this source is unresolved; no extended flux
is detected at the 0.2 mJy beam$^{-1}$ level.

1ES 1028+511:  This source is unresolved by both the VLBA (this paper) and in a VLA
snapshot by \citet{per96}.  The VLBA core flux is only $\sim 50$\% of the VLA core flux, indicating
either variability or extended flux which is too faint or resolved out on parsec scales. The redshift
for this source is tentative, and must be confirmed by further optical spectroscopy.



1ES 1212+078:  The VLBA map of this source (Figure~\ref{fig-12}) shows a well-collimated jet extending
to the east (PA = $+92$\arcdeg).  The jet is straight as far as 50 parsecs; and it has
several discrete components.  A deep VLA map of this source (Figure~\ref{fig-13}) shows 
an unusual, diffuse halo around the source, with no clear evidence of a jet in any direction.
Measuring from the core to the brightest hotspot gives PA = $+178$\arcdeg\ and \delpa\ $= 86$\arcdeg,
although this is highly speculative.  However, a very large \delpa\ does seem likely for this source.

1E 1415+259:  This source was detected by the VLBA, but due to its faintness its extended
structure cannot be modeled.  

RGB 1427+238:  The VLA map of this source (Figure~\ref{fig-14}) shows a compact structure that
consists of a core and either a halo or roughly collinear jets extending north PA $= -10$\arcdeg\ and
south PA $= -175$\arcdeg.  The VLBA source is unresolved on the 1.1 mJy beam$^{-1}$ level
\citep{fey00}.

1ES 1553+113:  The VLBA map (Figure~\ref{fig-15}) shows a jet extending to the northeast (PA =
+48\arcdeg).  Beyond $20$ parsecs the jet is very faint and diffuse, thus it is difficult to
determine the opening angle or if the jet is bent.
No jet continuous from the core is detected in a deep VLA map
(Figure~\ref{fig-16}); however a faint lobe is detected
south of the core, with a weak ``hot spot" at PA = +160\arcdeg, giving a large misalignment angle of
\delpa\
$= 112$\arcdeg\ if we assume the VLBA jet is related to the southern lobe.  

1ES 1741+196:  The VLBA map of this source (Figure~\ref{fig-17}) shows a well-collimated jet extending
to the east (PA = +86\arcdeg).  The jet is very straight, although it does show evidence
for 5\arcdeg\ bend to the south, 15--20 parsecs from the core.  
The VLA snapshot of this source \citep{per96} shows a jet which is well-aligned with the parsec-scale
jet (\delpa\ = 5\arcdeg).

\begin{figure}
\plotone{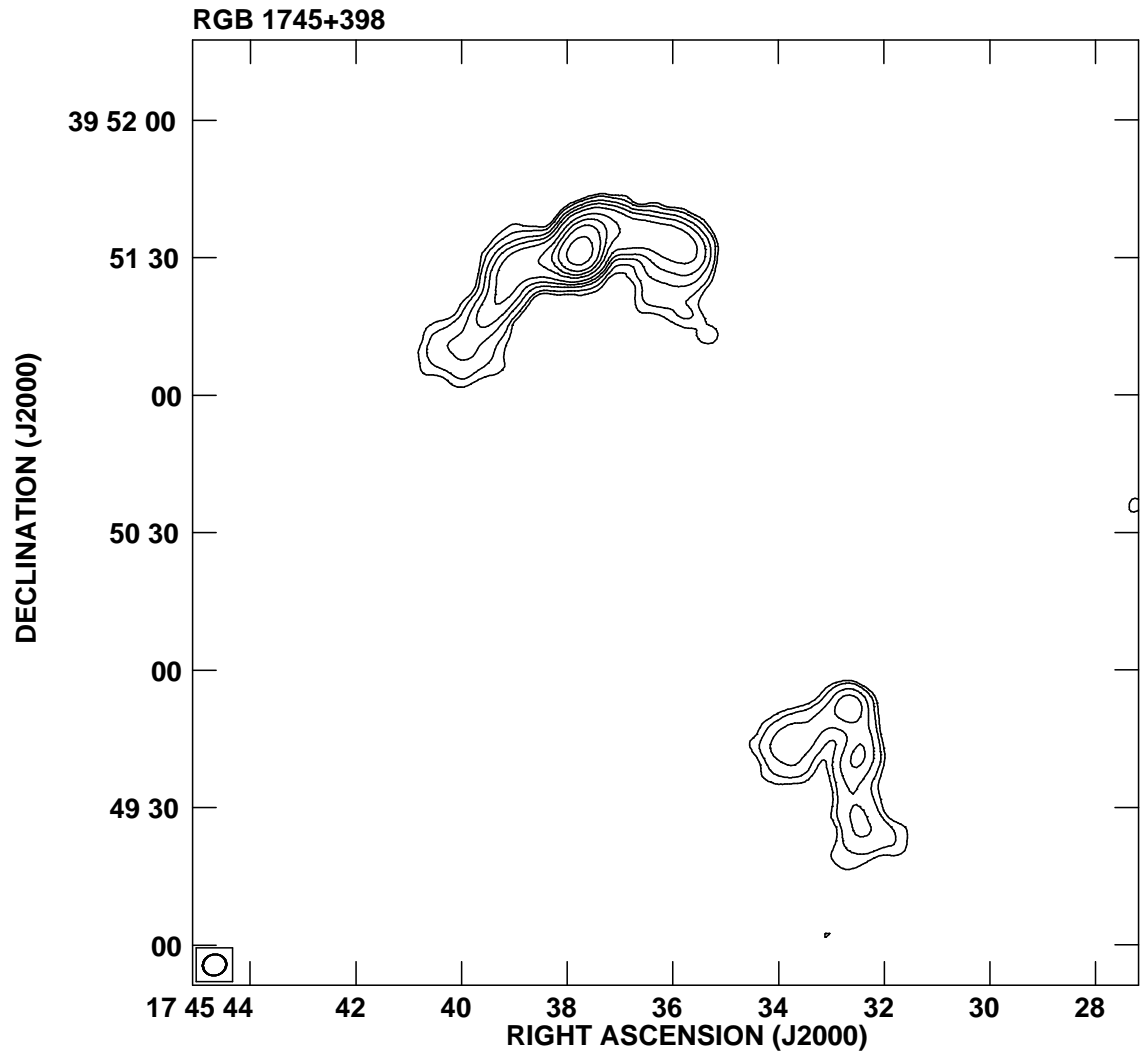}
\caption{VLA 1.425 GHz map of RGB 1745+398 (upper left) and a companion.  The beam is shown in the lower left
corner.  The contour levels are 0.1, 0.2, 0.5, 1, 2, 5, 10, 20, 50 and 100\% the peak flux of 3.159 x
10$^{-1}$ Jy beam$^{-1}$.}
\label{fig-18}
\end{figure}

\begin{figure}
\plotone{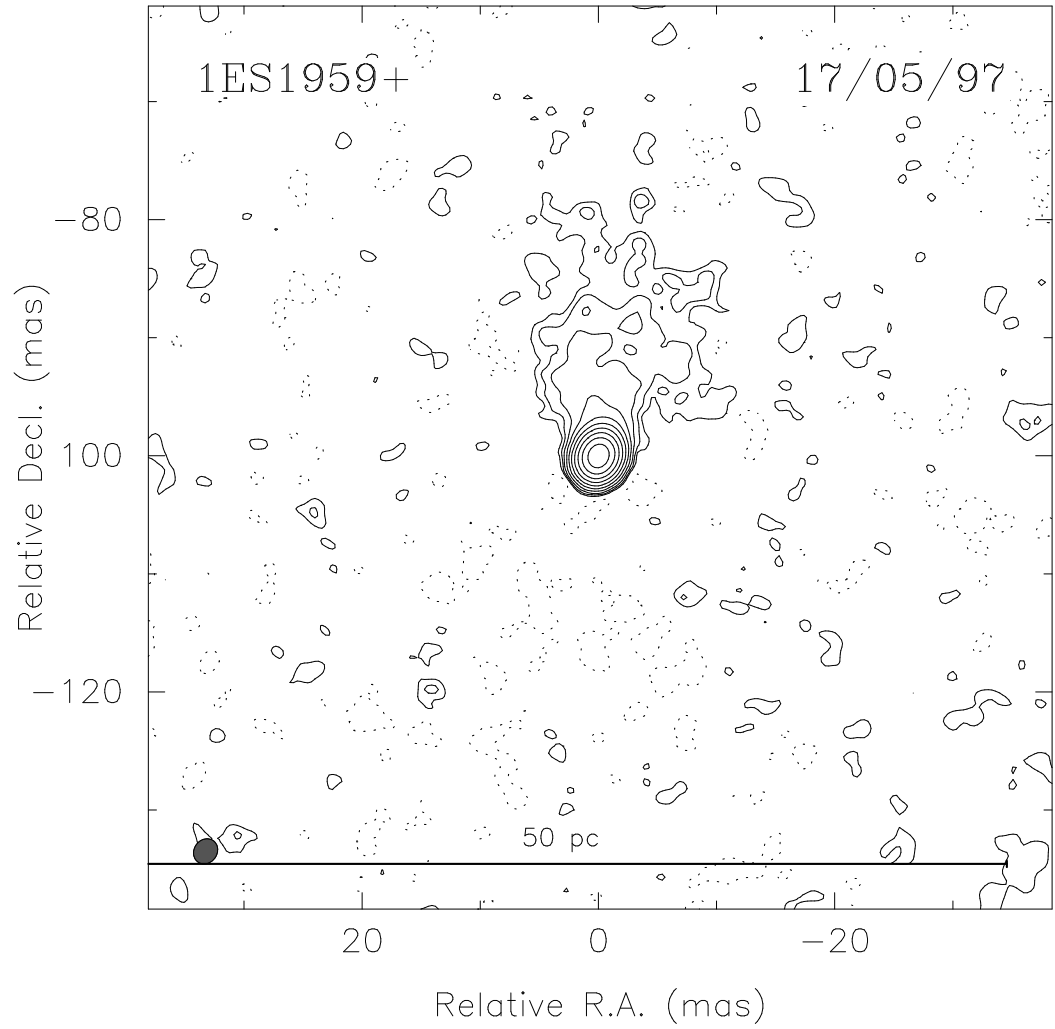}
\caption{VLBA 4.964 GHz map of 1ES 1959+650.  The contour levels are $-0.12$, 0.12, 0.23, 0.46, 0.92, 1.84,
3.69, 7.38, 14.76, 29.52 and 59.04\% of the peak flux of 1.734 x 10$^{-1}$ Jy beam$^{-1}$.  The beam, shown in
the lower left corner, has a FWHM of 2.23 x 1.94 mas, PA
$-35.9$\arcdeg.   The date of observation is shown in the upper right corner of the map.}
\label{fig-19}
\end{figure}

\begin{figure}
\plotone{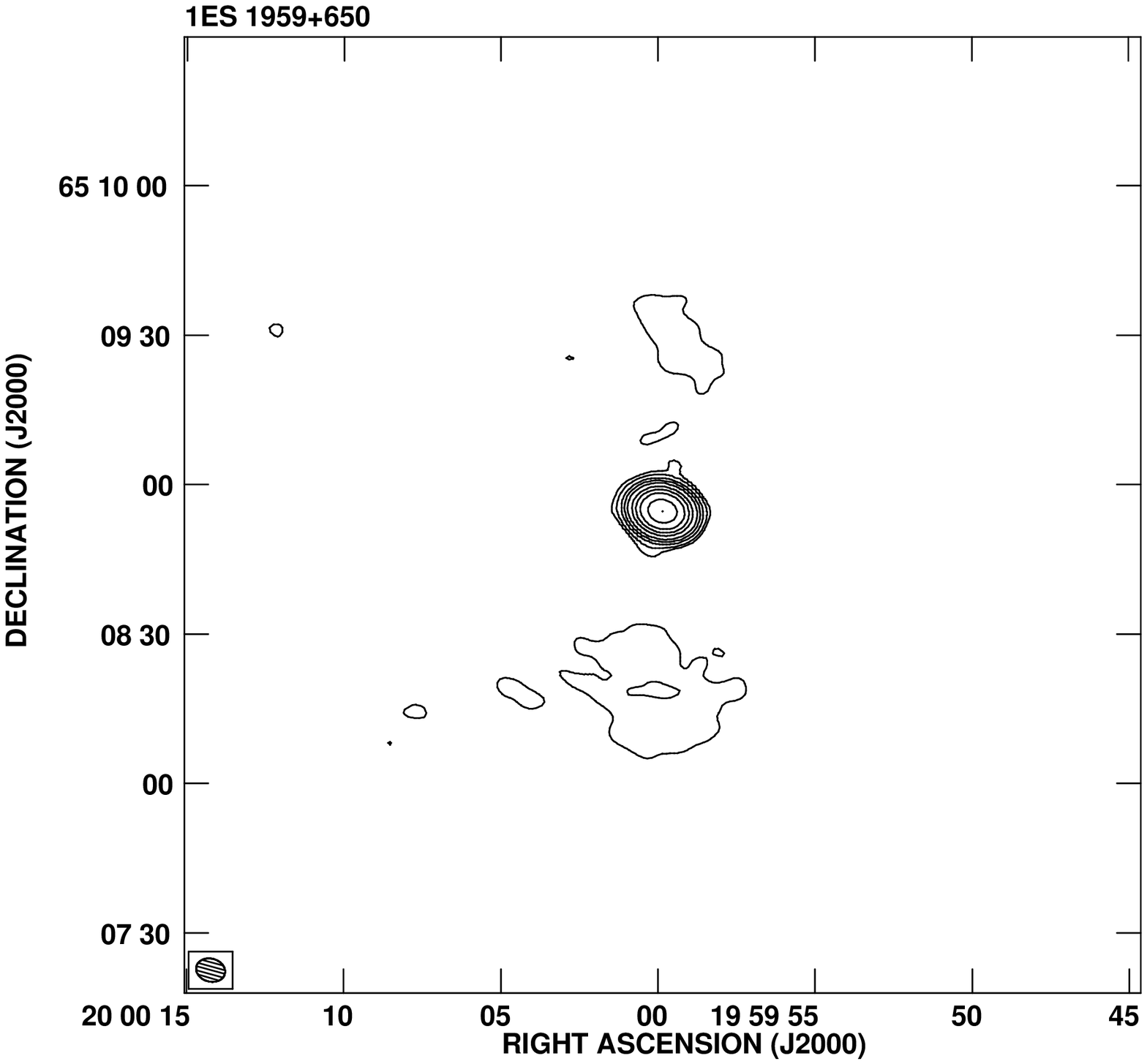}
\caption{VLA 1.425 GHz map of 1ES 1959+650.  The beam is shown in the lower left
corner.  The contour levels are 0.1, 0.2, 0.5, 1, 2, 5, 10, 20, 50 and 100\% the peak flux of 2.276 x
10$^{-1}$ Jy beam$^{-1}$.}
\label{fig-20}
\end{figure}

\begin{figure}
\plotone{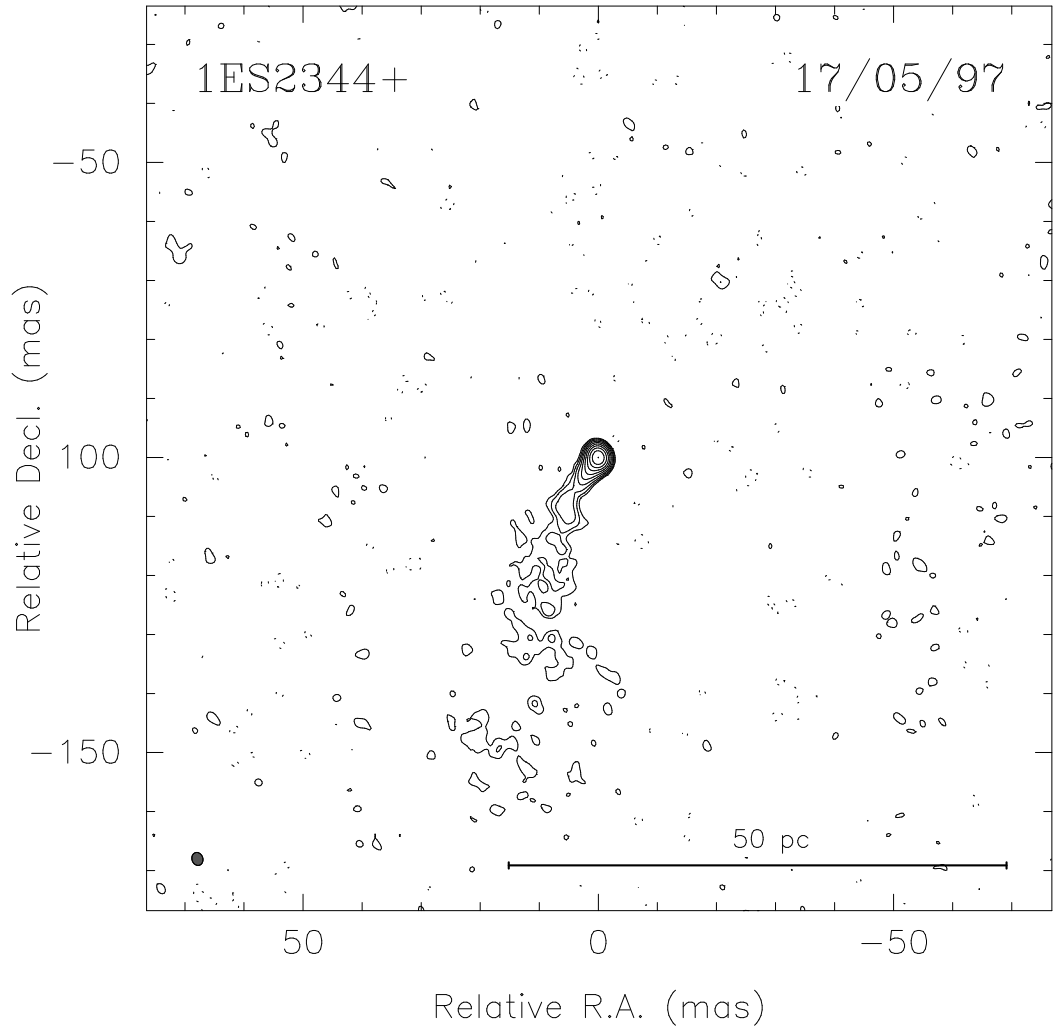}
\caption{VLBA 4.964 GHz map of 1ES 2344+514.  The contour levels are $-0.11$, 0.19, 0.39, 0.78, 1.55, 3.10,
6.21, 12.42, 24.84, 49.68 and 99.35\% of the peak flux of 1.031 x 10$^{-1}$ Jy beam$^{-1}$.  The beam, shown
in the lower left corner, has a FWHM of 2.23 x 1.94 mas, PA
$-35.9$\arcdeg.   The date of observation is shown in the upper right corner of the map.}
\label{fig-21}
\end{figure}

\begin{figure}
\plotone{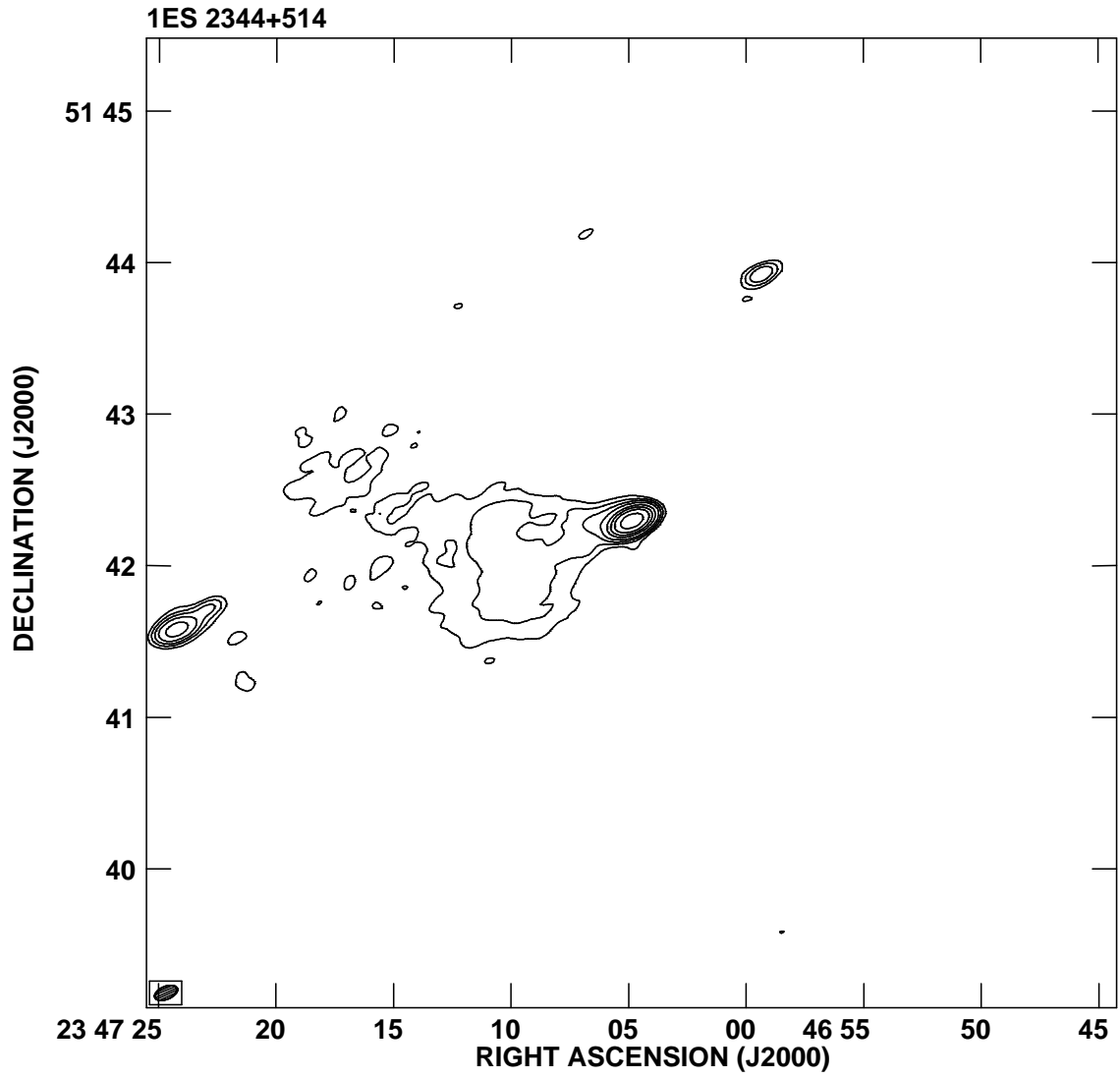}
\caption{VLA 1.425 GHz map of 1ES 2344+514.  The beam is shown in the lower left
corner.  The contour levels are 0.2, 0.5, 1, 2, 5, 10, 20, 50 and 100\% the peak flux of 2.108 x
10$^{-1}$ Jy beam$^{-1}$.}
\label{fig-22}
\end{figure}

RGB 1745+398:  RGB 1745+398 is the unusual object to the north in the VLA map shown in
Figure~\ref{fig-18}.  It has an edge-darkened, FR-I morphology that is highly distorted.  One jet appears to
emerge from the core at PA =
$-28$\arcdeg\ before bending to the west.  The other jet emerges from the core at PA $= +105$\arcdeg\ before
bending to the southeast.  Both jets show sharp bends, of up to 70\arcdeg.  This object lies close to the
center of a moderately massive galaxy cluster \citep{nil99}; thus its distorted shape may be the result of
interaction with either the ICM or the halos of other cluster members.  This hypothesis is supported by
the presence of another highly distorted radio source to the southwest of RGB 1745+398 in
Figure~\ref{fig-18}, which may also be a cluster member.  A VLBA map of RGB 1745+398 (Preeti,
private comm.) shows a component emerging due south from the core at PA $= -175$\arcdeg.  Thus
\delpa\ $= 80$\arcdeg\ in this very distorted source.

1ES 1959+650:  The VLBA map (Figure~\ref{fig-19}) shows a broad, diffuse jet to the north (PA $\approx
-5$\arcdeg).  The opening angle of the jet is wide ($\sim$55\arcdeg).  The core is
unresolved in a deep VLA map (Figure~\ref{fig-20}); however, very faint extended flux is detected to
the north (PA $\approx -5$\arcdeg) and south (PA $\approx +175$\arcdeg) of the core.  Thus the VLBA jet
appears to be well aligned with the northern lobe. 

1ES 2344+514:  The VLBA map (Figure~\ref{fig-21}) shows a jet extending to the southeast (PA = +145);
it appears to be well-collimated for about 10 parsecs before bending 25\arcdeg to the south
and broadening into a cone with a $\sim$35\arcdeg\ opening angle.  
A deep VLA map (Figure~\ref{fig-22}) detects emission extending to the east (PA = +105\arcdeg) in a
50\arcdeg\ cone.  The \delpa\ for this source is at least 40\arcdeg.  

\section{Results and Discussion}

Previous studies of the VLBI structure of BL Lac objects, nearly all of which were of LBLs, have shown
that large misalignment angles are common in these objects \citep{kol92,app96,cas02}.  This is to be
expected in sources that are seen close to the line of sight, as projection effects will magnify the
apparent distortion from intrinsic bends and complex structure within these jets.  

Measuring the parsec-scale jet PAs in BL Lacs is very difficult for several reasons.  Many of
these objects show jets which bend within several parsecs from the core; and in many cases the
emission where the bending occurs is very faint.  Thus the measured PA is very
sensitive to the linear resolution and the sensitivity of the observations, which of course depend on
the observed wavelength, the distance to the object, the ($u,v$) coverage and the overall quality of the
observations.  The parsec-scale jets seen in Mkn 501 and 1Jy 1147+245 are good examples of this problem
\citep{con95,gab99,cas02}.  
Adding another
dimension to the problem, there is evidence that the jet trajectory for some of these objects can change on short
timescales; e.g., 1Jy 0735+178 \citep{gom01}.   

Like the VLBI maps, the measured jet PAs in the VLA maps
are also dependent on sensitivity and resolution; e.g., the measured PAs can differ by as much as
60\arcdeg\ based upon the resolution of the maps for 1Jy 0814+425 and 1Jy 2131--021
\citep{cas02,rec02}.  Additionally, the kpc-scale structure of these sources are usually highly
distorted, often with a ``halo" that surrounds the core with no clear PA.  Thus, it is not surprising
that we see such a range of \delpa\ values, even without considering the enormous difference in
physical scales between the VLA and VLBI maps and the physical environments through which a jet
propagates from the core to kiloparsec scales.

Despite the uncertainties inherent in such measurements, we have measured parsec- and
kpc-scale jet position angles for our HBL sample.  
We include the four HBLs 
observed by \citet{kol96} into our HBL sample.
For comparison, we consider all of the LBLs in the complete 1Jy
sample \citep{sti91} for which high dynamic-range VLA maps exist \citep[and references
therein]{rec01,rec02} and that have been studied in detail with VLBI techniques
\citep{cas02,fey97,fey00,fom00,gab99,gab00b,gab00a,kel98,ros01,she97,she98}.  For consistency, we measure jet
position angles from these maps using the methodology described below rather than use published values.

There are only a few cases where the jet PA is unambiguous.  Thus the parsec- and kpc-scale jet position
angles were measured with the following methodology.  For both the VLA and VLBI maps the PA is measured from
the core through contiguous jet components that are more than 2--3$\sigma$ detections. In some cases the PA
on VLBI scales is uncertain because the jet is diffuse with a broad opening angle (e.g., 1ES 0806+524 and 1ES
1959+650).  In these cases the PA is measured either down the center of the jet or along the brightest
contours within the jet.  Also, there are four objects in our HBL sample (1ES 0033+595, 1ES 0414+009,
1ES 1741+196 and 1ES 2344+514) that have bright, well-collimated parsec-scale jets close to the core and show
evidence of bending further from the core.  However, the emission where the bending may be occuring is faint;
and it is not clear whether the jet is actually bending or simply broadening.  For this reason the jet PA
is measured from the bright jet components near the core.  In three of the four objects the kpc-scale
structure is well aligned with these measurements; the exception is 1ES 2344+514, for which no measurement of
the PA$_{\rm pc}$ will align it with the VLA PA$_{\rm kpc}$.  Thus it is possible that the jets in these objects
are bending and may not as well aligned as measured.  

The better alignment of the parsec-scale jet near the core with the kpc-scale jets in these sources may be
the result of collisions with dense clouds of gas near the core.  For a powerful jet, such collisions are not
effective at deflecting the jet in a coherent manner; however they may result in temporary distortions of the jets
on timescales of $< 10^7$ yr \citep{dey91,wan00}.  Thus, it is possible that the observed VLBI morphologies could
be explained by an off-center collision with a dense gas cloud 10--20 parsecs from the core which distorts the
observed PA at this distance but doesn't affect the long-term propagation of the jet to kpc
distances.

It is worth noting that \citet{hou02} find significant PA differences between inner and outer VLBI jet
components in a sample of lobe-dominated quasars, wherein the outer VLBI jet components are better aligned
with the kiloparsec-scale jets.  This suggests that, in quasars at least, jets may distort or bend close
to the core before becoming well-collimated further from the core.  The comparison is limited because quasar
jets are of much higher power; and due to the much higher redshift of their quasar sample ($z>1$), the VLBI
maps in \citet{hou02} are studying larger physical scales by a factor of $\sim$10.  Also, lobe-dominated
quasar jets are seen at large angles to the line of sight, and therefore do not suffer from projection
effects to nearly the degree as BL Lac objects.

In all of the resolved VLBI maps jet structure is apparent; however the VLA maps reveal that many of these
sources have a ``halo" morphology, i.e., extended emission that surrounds the core with no clear jet.  For
these objects the PA is measured to the brightest ``hotspot" within the lobe; and for objects which have two
distinct lobes and no clear jet, the PA is measured to the lobe most likely associated with the
parsec-scale jet (i.e., to the lobe which minimizes the value of \delpa).  Naturally,
\delpa\ could only be determined for objects which were resolved by both the VLA and with VLBI. 
%
%

\begin{figure}
\plotone{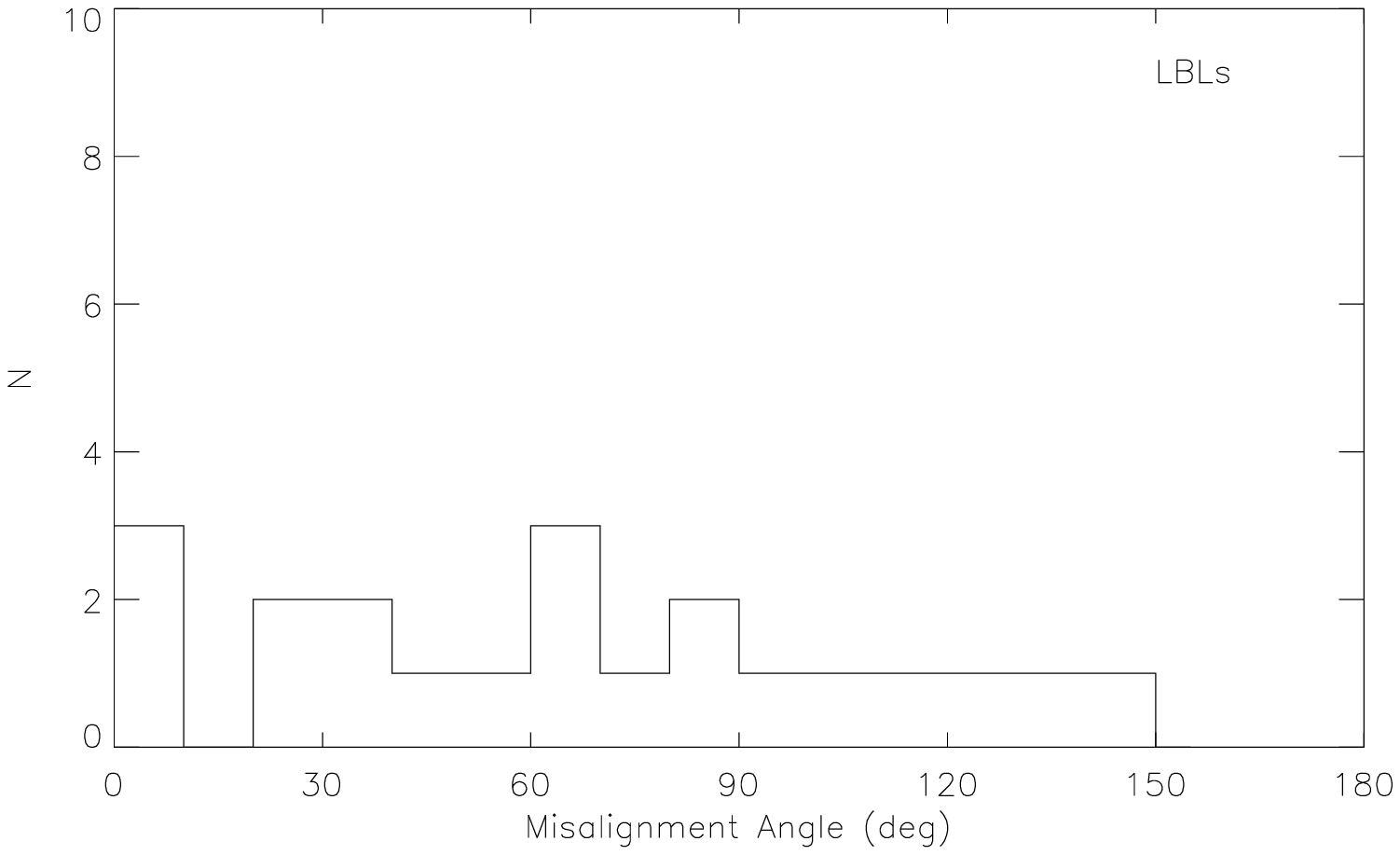}
\plotone{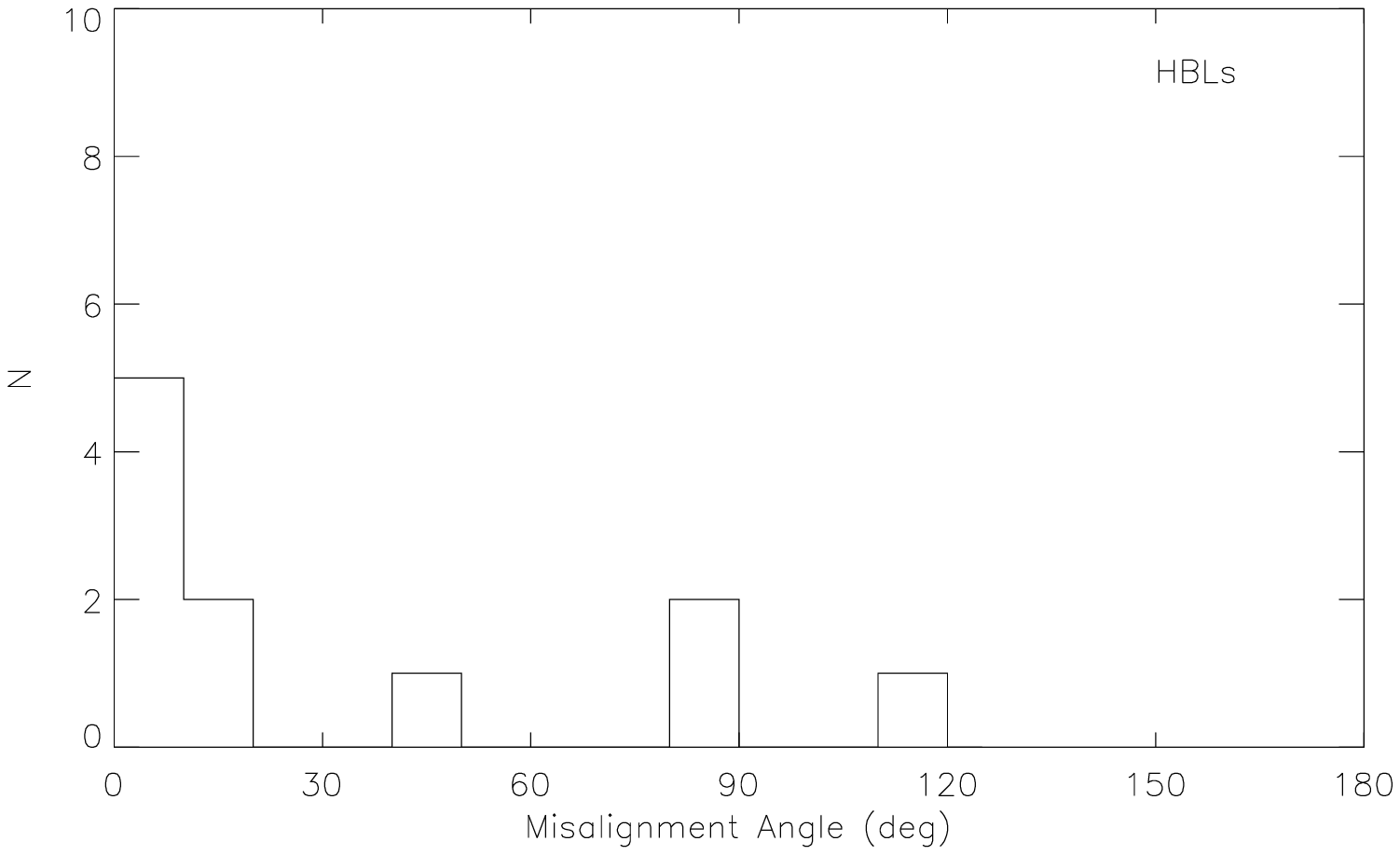}
\caption{Distribution of misalignment angle \delpa\ for LBLs from the 1Jy sample
(top) and HBLs from our sample (bottom).  The 1Jy LBL sample shows a
smooth distribution of misalignment angles from 0\arcdeg\ to 150\arcdeg, whereas
there is evidence that HBL jets are more well aligned.}
\label{fig-2}
\end{figure}

Figure~\ref{fig-2} shows the distribution of misalignment angles for the HBLs in our sample and the LBLs
in the 1Jy sample.     
Unfortunately, in only seven of the fifteen HBLs in our sample are jets resolved in both VLA and VLBI
images.  The parsec- and kpc-scale jets are well-aligned (\delpa\ $\leq 10$\arcdeg) for five of the
seven.  
In three additional sources a VLA PA value was determined by the brightest
hotspot in a lobe.  In only one of these cases is the parsec-scale jet well aligned with
the hotspot.  The remaining six sources are unresolved in one or both maps.  To this sample we added the
four HBLs in \citet{kol96}, although for only one could a \delpa\ value be
determined.  While all four are resolved in VLBI maps, a VLA map of 1ES 1133+704 by \citet{ulv86} shows
a halo morphology with no clear PA; and VLA maps of the other sources reliably resolve only 1ES 1727+503
\citep{ulv86,lau93}. In the 1Jy LBL sample we were
able to determine \delpa\ values for only 18 of 37 objects, with an additional three tentative \delpa\
values.  The difficulty in determining \delpa\ values in the 1Jy sample is due to unresolved sources
as well as many objects which show halo structures around the core with no resolved jet or hotspots.

Whereas most of the HBLs are well aligned (\delpa\ $\leq 20$\arcdeg), the LBL sample shows a wide
distribution of
\delpa\ values, evenly distributed from 0\arcdeg\ to 150\arcdeg.  The distributions of
\delpa\ for HBLs and LBLs are different at the 96\% level of confidence using a two-sided Kolmogorov-Smirnov
test.  Additionally, all four extreme HBLs (\logfxfr\ $> -4.5$) with well-measured misalignment angles
are very well aligned.  While not conclusive, this suggests that HBLs either have intriniscally straighter
jets than LBLs, or that HBLs are seen further off-axis than LBLs, such that projection effects are not as
important.  Clearly larger and statistically-complete samples are desirable.

We note that the distribution of \delpa\ values in the 1Jy LBL sample is consistent with the
distribution of \delpa\ values seen by \citet{kol92}.  We also note that there is no significant
evidence for a bimodal distribution of misalignment angles, with two peaks centered on
0\arcdeg\ and 90\arcdeg, as suggested for BL Lacs by \citet{app96}.  Thus we argue that the 1Jy LBL
sample is consistent with a population of radio galaxies whose jet axes are seen close to
the line of sight; and that complex bend geometries such as those proposed by \citet{con93}
are not necessary to explain the distributions of \delpa\ in either of our samples.

\begin{figure}
\plotone{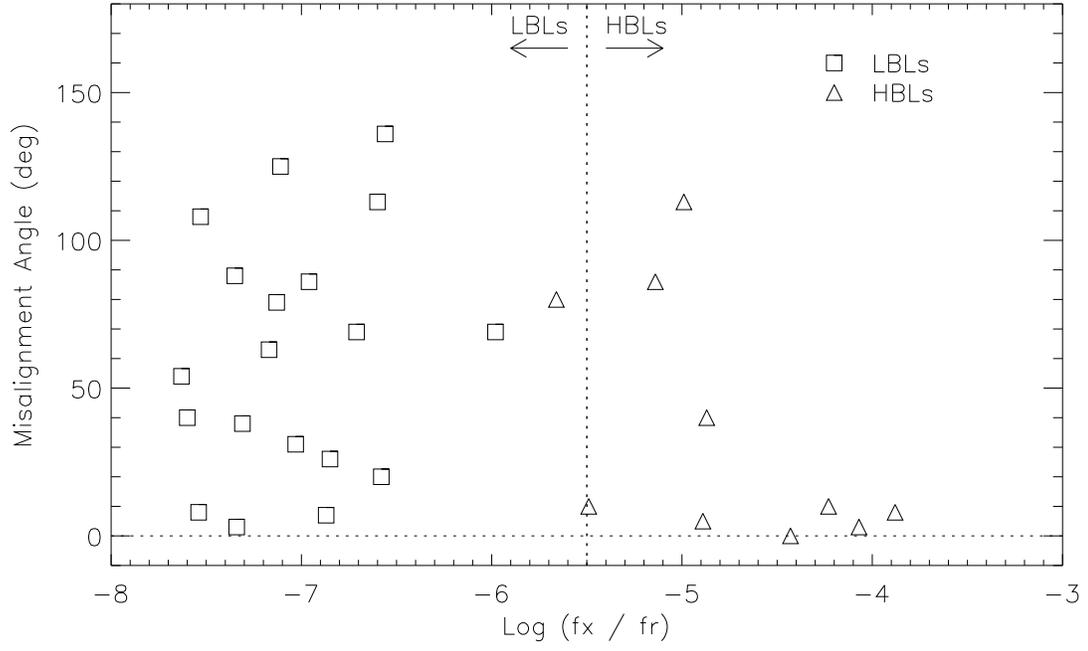}
\caption{Misalignment angle \delpa\ for LBLs (squares) and HBLs (triangles) as a
function of \logfxfr.  The dividing line at \logfxfr\ $\sim -5.5$ roughly divides HBLs and
LBLs.  }
\label{fig-3}
\end{figure}

If the orientation hypothesis is correct, that is, if HBLs are simply LBLs seen further
from the jet axis, there should be a correlation between \logfxfr\ and \delpa\ because LBLs will appear to be
more misaligned due to projection effects. Figure~\ref{fig-3} shows the distribution of \delpa\ as a function
of
\logfxfr\ for both samples.  There is a weak $2\sigma$ evidence (90\% probability) that \delpa\ and \logfxfr\
are linearly anticorrelated when the samples are combined; but this anticorrelation is not seen in the LBL
sample alone.  No correlation is seen between \delpa\ and $z$ for either the full LBL and HBL sample or
amongst the LBLs alone.

\section{Conclusions}

We have completed VLA and first-epoch VLBA observations as part of a program to study the
parsec-scale radio structure of a sample of fifteen HBLs.  
All of the resolved objects are core dominated in the VLA and VLBA maps; and they show a
core-jet morphology on parsec scales, similar to LBLs and core-dominated quasars.  Some sources show a
well-collimated, parsec-scale jet with discrete components, whereas others show a diffuse jet with a
wide opening angle.  Some objects also show evidence for the interaction of their parsec-scale jets with
a dense gas environment.  Further modeling is warranted.

While LBLs show a wide distribution of parsec- and
kpc-scale jet alignment angles, most of the HBLs considered here have well-aligned jets,
suggesting either that HBL jets are seen further off-axis than LBL jets, or that HBL
jets are intrinsically straighter.  
Complex bend geometries, such as those proposed by \citet{con93}, are not necessary to explain the
observed distributions of misalignment angles seen in our LBL or HBL samples.  
There is a hint in our data that extreme HBLs (\logfxfr\ $> -4.5$) have intrinsically straight jets that are
viewed well off-axis in that all four extreme HBLs in our sample have very small ($< 10$\%) misalignment
angles.  Observations of other extreme HBLs are needed to test this preliminary result.

 
While it is now clear that orientation alone cannot be invoked to unify HBLs and LBLs \citep{rec01},
selection effects may nonetheless cause LBLs to be seen closer to the jet axis than HBLs.  For
example, LBLs show optical emission lines which are several orders of magnitude more luminous than
HBLs.  Thus LBLs may require larger Doppler factors to sufficiently boost the jet continuum relative to
the emission lines to remain within the BL Lac spectral criterion (rest $W_{\lambda} \leq 5$\AA). 
Similarly, the bright radio flux limit as well as the flat radio spectrum criterion of the 1Jy sample may
also bias LBLs towards more highly beamed objects, whereas the X-ray surveys used to draw our HBL sample do
not have any radio-based selection criteria.  Thus, LBLs may be systematically more beamed than HBLs even
though they may not necessarily share the same parent population.



\acknowledgments

Research on BL Lac objects at the University of Colorado was supported by NASA grant
NAGW-2675.  Part of this work is a part of a Ph.D. dissertation submitted to the University of Colorado
by T.A.R.

\clearpage


%
%




\clearpage

\begin{deluxetable}{rllcl}
\tablecaption{General Properties of the Sample \label{tbl-1}}
\tablewidth{0pt}
\tablehead{
\colhead{Object} & \colhead{RA(J2000)} & \colhead{Dec} & \colhead{ \logfxfr } & \colhead{$z$}
}
\startdata
1ES 0033+595 & 00:35:52.644 & +59:50:04.59 & -4.07 & 0.086  \\
1ES 0229+200 & 02:32:48.616 & +20:17:17.45 & -4.23 & 0.139  \\
1ES 0414+009 & 04:16:52.494 & +01:05:23.91 & -3.88 & 0.287  \\
1ES 0647+250 & 06:50:46.489 & +25:02:59.63 & -4.09 & 0.203: \\ 
RGB 0656+426 & 06:56:10.72  & +42:37:02.7  & -5.49 & 0.059 \\
1ES 0806+524 & 08:09:49.188 & +52:18:58.24 & -4.69 & 0.136  \\ 
1ES 1028+511 & 10:31:18.524 & +50:53:35.79 & -3.96 & 0.359: \\ 
1ES 1212+078 & 12:15:10.977 & +07:32:04.67 & -5.14 & 0.135  \\ 
1E  1415+259 & 14:17:56.680\tablenotemark{a} &  +25:43:26.24 & -4.56 & 0.237 \\ 
RGB 1427+238 & 14:27:00.392\tablenotemark{b} & +23:48:00.04 & -5.60 & \nodata \\
1ES 1553+113 & 15:55:43.044 & +11:11:24.37 & -4.99 & 0.360 \\
1ES 1741+196 & 17:43:57.838 & +19:35:08.99 & -4.89 & 0.083 \\
RGB 1745+398 & 17:45:37.71  & +39:51:31.8  & -5.66 & 0.267 \\
1ES 1959+650 & 19:59:59.852 & +65:08:54.69 & -4.43 & 0.048 \\ 
1ES 2344+514 & 23:47:04.838 & +51:42:17.88 & -4.87 & 0.044 \\
\enddata
\tablenotetext{a}{The core position for 1E 1415+259 was determined from the FIRST survey.}
\tablenotetext{b}{The core position for RGB 1427+238 was reported in \citet{ma98}.}
\end{deluxetable}

\begin{deluxetable}{rrrrrrr}
\tablecaption{Radio Properties of the Sample \label{tbl-2}}
\tablewidth{0pt}
\tablehead{
\colhead{} & \multicolumn{3}{c}{VLA (1.4 GHz)} &\multicolumn{3}{c}{VLBA (6cm)}  \\
\cline{2-4} \cline{5-7} \\
\colhead{Object} & \colhead{$f_{core}$ (mJy)} & \colhead{$f_{ext}$ (mJy)} & \colhead{PA$_{\rm
kpc}$\tablenotemark{a}} &
\colhead{$f_{core}$ (mJy)} & \colhead{$f_{ext}$ (mJy)} & \colhead{PA$_{\rm pc}$\tablenotemark{a}} 
}
\startdata
1ES	0033+595	&	90.4	   &	$61.3\pm12$  	&	+62\arcdeg &	39.7   	&	$12.8\pm0.8$	&	+65\arcdeg  \\	
1ES	0229+200	&	51.8   	&	$34.7\pm8.7$	 &	180\arcdeg &	22.7   	&	$7.7\pm0.8$ 	&	+170\arcdeg \\	
1ES	0414+009	&	\nodata	&	\nodata	      &	+63\arcdeg &	36.4   	&	$12.9\pm0.8$	&	+68\arcdeg  \\	
1ES	0647+250	&	\nodata	&	\nodata	      &	\nodata    &	46.0   	&	$11.0\pm0.7$	&	\nodata     \\	
RGB	0656+426	&	252.8	  &	$699\pm32$   	&	$-140$\arcdeg &	\nodata	&	\nodata	     &	$-150$\arcdeg \\	
1ES	0806+524	&	189.1  	&	$<10$        	&	\nodata    &	111.2	  &	$39.5\pm1.1$	&	+13\arcdeg \\	
1ES	1028+511	&	215.7	  &	$<15$	        &	\nodata    &	22.4	   &	$3.5\pm0.4$ 	&	\nodata    \\	
1ES	1212+078	&	87.4	   &	$48.6\pm12.9$	&	+178\arcdeg &	36.3   	&	$16.6\pm0.8$	&	+92\arcdeg \\	
1E	 1415+259	&	\nodata	&	\nodata      	&	\nodata     &	8.7	    &	$17.1\pm0.7$	&	\nodata    \\	
RGB	1427+238	&	331.7	  &	$130.2\pm2.9$	&	$-10$\arcdeg &	\nodata	&	\nodata	     &	\nodata	    \\	
1ES	1553+113	&	271.1  	&	$13.9\pm4.8$ 	&	+160\arcdeg &	258.7  	&	$21.8\pm0.6$	&	+48\arcdeg \\	
1ES	1741+196	&	\nodata	&	\nodata      	&	+91\arcdeg &	83.5	   &	$38.2\pm0.9$	&	+86\arcdeg \\	
RGB	1745+398	&	372.5	  &	$272.2\pm5.5$	&	+105\arcdeg &	\nodata	&	\nodata     	&		$-175$\arcdeg \\ 
1ES	1959+650	&	227.5  	&	$18\pm10$	    &	$-5$\arcdeg &	181.7	  &	$37.7\pm1.0$	&	 $-5$\arcdeg \\	
1ES	2344+514	&	217.0   &	$128\pm50$	   &	+105\arcdeg &	109.   	&	$59.7\pm2.0$	&	+145\arcdeg \\	
\enddata
\tablenotetext{a}{Please read \S 2.4 and \S 3 regarding the measurement of jet PAs.}
\end{deluxetable}

\clearpage

\clearpage

\clearpage

\clearpage

\clearpage

\clearpage

\clearpage

\clearpage

\clearpage

\clearpage

\clearpage

\clearpage

\clearpage

\clearpage

\clearpage

\clearpage

\clearpage

\clearpage

\clearpage

\clearpage

\clearpage

\clearpage





\begin{thebibliography}{}
\bibitem[Auri\`ere(1982)]{aur82} Auri\`ere, M.  1982, \aap\ 109, 301.
\bibitem[Abraham et al.(1991)]{abr91} Abraham, R.G., Crawford, C.S. \& McHardy, I.M. \mnras\ 1991
    252, 482.
\bibitem[Appl et al.(1996)]{app96} Appl, S., Sol, H. \& Vicente, L. 1996 \aap\ 310, 419.
\bibitem[Antonucci \& Ulvestad (1985)]{ant85} Antonucci, R.R.J. \& Ulvestad, J.S. 1985 \apj\ 294, 158.
\bibitem[Bondi et al. (2001)]{bon01} Bondi, M., March\~a, M.J.M., Dallacasa, D. \& Stanghellini, C.
2001 \mnras\ 325, 1109.
\bibitem[Briggs(1995)]{bri95} Briggs, D. 1995, PhD dissertation, New Mexico Institute of Mining and
Technology.
\bibitem[Brinkmann et al.(1996)]{bri96} Brinkmann, W., Siebert, J., Kollgaard, R.I. \& Thomas, H.-C.
    1996 \aap\ 313, 356.
\bibitem[Cassaro et al.(2002)]{cas02} Cassaro, P., Stanghellini, C., Dallacasa, D., Bondi, M. \&
Zappal\`a, R.A. 2002 \aap\ 381, 378.
\bibitem[Celotti et al.(1993)]{cel93} Celotti, A., Maraschi, L., Ghisellini, G., Caccianiga, A. \&
Maccacaro, T. 1993
\apj\ 416, 118.
\bibitem[Conway \& Murphy(1993)]{con93} Conway, J.E. \& Murphy, D.W. 1993 \apj\ 411, 89.
\bibitem[Conway \& Wrobel(1995)]{con95} Conway, J.E. \& Wrobel, J.M. 1995 \apj\ 439, 98.
\bibitem[Cohen(1988)]{coh88} Cohen, M.H. 1988, in BL Lac Objects, ed. L. Maraschi, T. Maccacaro \&
M.-H. Ulrich (Heidelberg: Springer-Verlag), 13.
\bibitem[De Young(1991)]{dey91} De Young, D.S. 1991 \apj\ 371, 69.
\bibitem[Fanaroff \& Riley(1974)]{fan74} Fanaroff, B.L. \& Riley, J.M. 1974 \mnras\ 167, 31.
\bibitem[Fey \& Charlot(1997)]{fey97} Fey, A.L. \& Charlot, P. 1997 \apjs\ 111, 95.
\bibitem[Fey \& Charlot(2000)]{fey00} Fey, A.L. \& Charlot, P. 2000 \apjs\ 128, 17.
\bibitem[Fleming et al.(1993)]{fle93} Fleming, T.A., Green, R.F., Jannuzi, B.T.,
Liebert, J., Smith, P.S. \& Fink, H. 1993 \aj\ 106, 1729.
\bibitem[Fomalont et al.(2000)]{fom00} Fomalont, E.B., Frey, S., Paragi, Z., Gurvits, L.I. \& Scott,
W.K., Taylor, A.R., Edwards, P.G. \& Hirabayshi, H. 2000 \apjs\ 131, 95.
\bibitem[Fossati et al.(1998)]{fos98} Fossati, G., Maraschi, L., Celotti, A., Comastri, A. \&
Ghisellini, G. 1998 \mnras\ 299, 433. 
\bibitem[Gabuzda et al.(1999)]{gab99} Gabuzda, D.C., Pushkarev, A.B. \& Cawthorne, T.V. 1999 \mnras\
307, 725. 
\bibitem[Gabuzda \& Cawthorne(2000)]{gab00a} Gabuzda, D.C. \& Cawthorne, T.V. 2000 \mnras\ 319, 1056. 
\bibitem[Gabuzda et al.(2000)]{gab00b} Gabuzda, D.C., Pushkarev, A.B. \& Cawthorne, T.V. 2000 \mnras\
319, 1109. 
\bibitem[Ghisellini et al.(1993)]{ghi93} Ghisellini, G., Padovani, P. Celotti, A. \& Maraschi, L.
1993 \apj\ 407, 65.
\bibitem[Giommi \& Padovani(1994)]{gio94} Giommi, P. \& Padovani, P. 1994 \mnras\ 268, L51.
\bibitem[G\'omez et al.(2001)]{gom01} G\'omez, J.L., Guirado, J.C., Agudo, I., Marscher, A.P., Alberdi,
A., Marcaide, J.M. \& Gabuzda, D.C. 2001 \mnras\ 328, 873.
\bibitem[Hough et al.(2002)]{hou02} Hough, D.H., Vermeulen, R.C., Readhead, A.C.S., Cross, L.L., Barth, E.L.,
Yu, L.H., Beyer, P.J. \& Phifer, E.M. 2002 \aj\ 123, 1258.
\bibitem[Jannuzi(1990)]{jan90} Jannuzi, B.T. 1990 PhD thesis, U. of Arizona.
\bibitem[Jannuzi et al.(1993)]{jan93} Jannuzi, B.T., Smith, P.S. \& Elston, R. 1993 \apjs\ 85, 265.
\bibitem[Jannuzi et al.(1994)]{jan94} Jannuzi, B.T., Elston, R. and Smith, P. 1994 \apj\ 428, 130.
\bibitem[Kellermann et al.(1998)]{kel98} Kellermann, K.I., Vermeulen, R.C., Zensus, J.A. \& Cohen, M.H.
1998 \apj\ 115, 1295.
\bibitem[Kollgaard et al.(1992)]{kol92}Kollgaard, R.I., Wardle, J.F.C., Roberts, D.H. \& 
Gabuzda, D.C. 1992 \aj\ 104, 1687.
\bibitem[Kollgaard et al.(1996)]{kol96} Kollgaard, R.I., Gabuzda, D.C. and Feigelson, E.D. 1996
\apj\ 460, 164.
\bibitem[Laurent-Muehleisen et al.(1993)]{lau93} Laurent-Muehleisen, S.A., Kollgaard, R.I.,
Moellenbrock, G.A. \& Feigelson, E.D. 1993 \aj\ 106, 875.
\bibitem[Laurent-Muehleisen et al.(1999)]{lau99} Laurent-Muehleisen, S.A., Kollgaard, R.I.,
Feigelson, E.D., Brinkmann, W. \& Siebert, J. 1999 \apj\ 525, 127.
\bibitem[Ma et al.(1998)]{ma98} Ma, C., Arias, E.F., Eubanks, T.M., Fey, A.L., Gontier, A.-M., Jacobs,
C.S., Sovers, O.J., Archinal, B.A. \& Charlot, P. 1998 \aj\ 116, 516.
\bibitem[Morris et al.(1991)]{mor91} Morris, S.L., Stocke, J.T., Gioia, I.M., Schild, R.E., Wolter,
A. \& Della Ceca, R. 1991 \apj\ 380, 49.
\bibitem[Nilsson et al.(1999)]{nil99} Nilsson, K., Takalo, L.O., Pursimo, T., Sillanp\"a\"a, A.,
Heidt, J., Wagner, S.J., Laurent-Muehleisen, S.A. \& Brinkmann, W. \aap\ 343, 81.
\bibitem[O'Dea et al.(1992)]{ode92} O'Dea, C.P., Baum, S.A., Stanghellini, C., Dey, A., van Breugel, W., Deustua,
S. \& Smith, E.P. 1992 \aj\ 104, 1320.
\bibitem[Orr \& Browne(1982)]{orr82} Orr, M.J.L. \& Browne, I.W.A. 1982 \mnras\ 200, 1067.
\bibitem[Padovani \& Urry(1990)]{pad90} Padovani, P. \& Urry, C.M. 1990 \apj\ 356, 75.
\bibitem[Padovani \& Giommi(1995)]{pad95} Padovani, P. \& Giommi, P. 1995 \apj\ 444, 567.
\bibitem[Pearson \& Readhead(1988)]{pea88} Pearson, T.J. \& Readhead, A.C.S. 1988 \apj\ 328, 114.
\bibitem[Perlman \& Stocke(1993)]{per93} Perlman, E.S. \& Stocke, J.T. 1993 \apj\ 406, 430.
\bibitem[Perlman \& Stocke(1994)]{per94} Perlman, E.S. \& Stocke, J.T. 1994 \aj\ 108, 56.
\bibitem[Perlman et al.(1996)]{per96} Perlman, E.S., Stocke, J.T., Schachter, J.F., Elvis, M.,
Ellingson, E., Urry, C.M., Potter, M., Impey, C.D. \& Kolchinsky, P. 1996 \apjs\ 104, 251.  
\bibitem[Perlman et al.(1998)]{per98} Perlman, E.S., Padovani, P., Giommi, P., Sambruna, R., Jones, L.R.,
Tzioumis, A. \& Reynolds, J. 1998 AJ 115, 1253.
\bibitem[Perlman et al.(1999)]{per99} Perlman, E.S., Schachter, J.F. \& Stocke, J.T. 1999, in prep.
\bibitem[Rector et al.(2000)]{rec00} Rector, T.A., Stocke, J.T., Perlman, E.S., Morris, S.L. \&
Gioia, I.A. 2000 \aj\ 120, 1626.
\bibitem[Rector \& Stocke(2001)]{rec01} Rector, T.A. \& Stocke, J.T. 2001 \aj\ 122, 565.
\bibitem[Rector \& Stocke(2002)]{rec02} Rector, T.A. \& Stocke, J.T. 2002, in prep.
\bibitem[Ros et al.(2001)]{ros01} Ros, E., Marcaide, J.M., Guirado, J.C. \& P\'erez-Torres, M.A.
2001 \aap\ 376, 1105.
\bibitem[Sambruna et al.(1996)]{sam96} Sambruna, R.M. Maraschi, L. \& Urry, M. 1996 \apj\ 463, 444.
\bibitem[Scarpa et al.(1999)]{sca99} Scarpa, R., Urry, C.M., Falomo, R., Pesce, J.E., Webster, R., O'Dowd, M.
\& Treves, A. 1999 \apj\ 521, 134.
\bibitem[Schwartz et al.(1989)]{sch89} Schwartz, D.A., Brissenden, R.J.V., Tuohy, I.R., Feigelson,
E.D., Hertz, P.L. \& Remillard, R.A. 1989 BAAS 21, 777.
\bibitem[Shen et al.(1997)]{she97} Shen, Z.-Q., Wan, T.-S., Moran, J.M., Jauncey, D.L., Reynolds, J.E.,
Tzioumis, A.K., Gough, R.G., Ferris, R.H., Sinclair, M.W., Jiang, D.-R., Hong, X.-Y., Liang, S.-G.,
Costa, M.E., Tingay, S.J., McCulloch, P.M., Lovell, J.E.J., King, E.A., Nicolson, G.D., Murphy, D.W.,
Meier, D.L., van Ommen, T.D. \& White, G.L. 1997 \aj\ 114, 1999.
\bibitem[Shen et al.(1998)]{she98} Shen, Z.-Q., Wan, T.-S., Moran, J.M., Jauncey, D.L., Reynolds, J.E.,
Tzioumis, A.K., Gough, R.G., Ferris, R.H., Sinclair, M.W., Jiang, D.-R., Hong, X.-Y., Liang, S.-G.,
Edwards, P.G., Costa, M.E., Tingay, S.J., McCulloch, P.M., Lovell, J.E.J., King, E.A., Nicolson, G.D.,
Murphy, D.W., Meier, D.L., van Ommen, T.D., Edwards, P.G. \& White, G.L. 1998 \aj\ 1115, 1357.
\bibitem[Shepherd(1997)]{shep97} Shepherd, M.C. 1997, in ASP Conf. Ser. 125, Astronomical Data Analysis
Software and Systems IV, eds. G. Hunt \& H.F. Payne (San Francisco: ASP), 77.
\bibitem[Stickel et al.(1991)]{sti91} Stickel, M., Padovani, P., Urry, C.M., Fried, J.W. \& K\"uhr,
H. 1991 \apj\ 374, 431.
\bibitem[Stocke(1989)]{sto89} Stocke, J.T. 1989, in BL Lac Objects, ed. L. Maraschi, T. Maccacaro \&
M.-H. Ulrich (Heidelberg: Springer-Verlag), 242.
\bibitem[Stocke et al.(1991)]{sto91} Stocke, J.T., Morris, S.L., Gioia, I.M., Maccacaro, T.,
Schild, R., Wolter, A., Fleming, T.A. \& Henry, J.P. 1991 \apjs\ 76, 813.
\bibitem[Ulvestad \& Antonucci(1986)]{ulv86} Ulvestad, J.S. \& Antonucci, R.R.J. 1986 \aj\ 92, 6.
\bibitem[Urry et al.(1991)]{urr91} Urry, C.M., Padovani, P. \& Stickel, M. 1991 \apj\ 382, 501.
\bibitem[Urry \& Padovani(1995)]{urr95} Urry, C.M. \& Padovani, P. 1995 \pasp\ 107, 803.
\bibitem[Wang, Witta \& Hooda(2000)]{wan00} Wang, Z., Wiita, P.J. \& Hooda, J.S. 2000 \apj\ 534, 201.
\bibitem[Wardle et al.(1984)]{war84} Wardle, J.F.C., Moore, R.L. \& Angel, J.R.P. 1984 \apj\ 279, 93.
\bibitem[Wurtz et al.(1996)]{wur96} Wurtz, R., Stocke, J.T. \& Yee, H.K.C. 1996 \apjs\ 103, 109.

\end{thebibliography}
\end{document}